\documentclass[proof]{pasj01}
\usepackage{natbib}
\Received{$\langle$reception date$\rangle$}
\Accepted{$\langle$acception date$\rangle$}
\Published{$\langle$publication date$\rangle$}

\begin{document}

\title{Cosmic-ray-driven enhancement of the C$^0$/CO abundance ratio in W51C}

\author{Mitsuyoshi Yamagishi\altaffilmark{1}}
\altaffiltext{1}{Institute of Astronomy, The University of Tokyo, 2-21-1 Osawa, Mitaka, Tokyo 181-0015, Japan}
\email{yamagishi@ioa.s.u-tokyo.ac.jp}

\author{Kenji Furuya\altaffilmark{2}}
\altaffiltext{2}{National Astronomical Observatory of Japan, National Institutes of Natural Sciences, 2-21-1 Osawa, Mitaka, Tokyo 181-8588, Japan}

\author{Hidetoshi Sano\altaffilmark{3}}
\altaffiltext{3}{Faculty of Engineering, Gifu University, 1-1 Yanagido, Gifu 501-1193, Japan}

\author{Natsuko Izumi\altaffilmark{4}}
\altaffiltext{4}{Institute of Astronomy and Astrophysics, Academia Sinica, No. 1, Section 4, Roosevelt Road, Taipei 10617, Taiwan}

\author{Tatsuya Takekoshi\altaffilmark{5}}
\altaffiltext{5}{Kitami Institute of Technology, 165 Koen-cho, Kitami, Hokkaido 090-8507, Japan}

\author{Hidehiro Kaneda\altaffilmark{6}}
\altaffiltext{6}{Graduate School of Science, Nagoya University, Chikusa-ku, 464-8602 Nagoya, Japan}

\author{Kouichiro Nakanishi\altaffilmark{2,7}}
\altaffiltext{7}{Department of Astronomical Science, The Graduate University for Advanced Studies, SOKENDAI, 2-21-1 Osawa, Mitaka, Tokyo 181-8588, Japan}

\author{Takashi Shimonishi\altaffilmark{8}}
\altaffiltext{8}{Institute of Science and Technology, Niigata University, Ikarashi-nihoncho 8050, Nishi-ku, Niigata 950-2181, Japan}

\KeyWords{ISM: individual objects (W51C) --- ISM: atoms --- cosmic rays ---  submillimeter: ISM}

\maketitle

\begin{abstract}
We examine spatial variations of the C$^0$/CO abundance ratio ($X_{\mathrm{C/CO}}$) in the vicinity of the $\gamma$-ray supernova remnant W51C, based on [C$\;${\sc i}] ($^3P_1$--$^3P_0$), $^{12}$CO(1--0), and $^{13}$CO(1--0) observations with the ASTE and Nobeyama 45-m telescopes.
We find that $X_{\mathrm{C/CO}}$ varies in a range of 0.02--0.16 (0.05 in median) inside the molecular clouds of $A_V>$100~mag, where photodissociation of CO by the interstellar UV is negligible.
Furthermore, $X_{\mathrm{C/CO}}$ is locally enhanced up to by a factor of four near the W51C center, depending on the projected distance from the W51C center.
In high-$A_V$ molecular clouds, $X_{\mathrm{C/CO}}$ is determined by the ratio of the cosmic-ray (CR) ionization rate to the H$_2$ density, and we find no clear spatial variation of the H$_2$ density against the projected distance. 
Hence, the high CR ionization rate may locally enhance $X_{\mathrm{C/CO}}$ near the W51C center.
We also find that the observed spatial extent of the enhanced $X_{\mathrm{C/CO}}$ ($\sim$17~pc) is consistent with the diffusion distance of CRs with the energy of 100~MeV.
The fact suggests that the low-energy CRs accelerated in W51C enhance $X_{\mathrm{C/CO}}$.
The CR ionization rate at the $X_{\mathrm{C/CO}}$-enhanced cloud is estimated to be 3$\times$10$^{-16}$~s$^{-1}$ on the basis of time-dependent PDR simulations of $X_{\mathrm{C/CO}}$, the value of which is 30 times higher than that in the standard Galactic environment.
These results demonstrate that [C$\;${\sc i}] is a powerful probe to investigate the interaction between CRs and the interstellar medium for a wide area in the vicinity of supernova remnants.
\end{abstract}

\section{Introduction}

Cosmic rays (CRs) are one of the important components that determine the chemical properties of the interstellar medium (ISM).
CRs penetrate deep into molecular clouds ($A_{\mathrm V}>$5~mag; e.g., \citealt{Shen04}), while interstellar ultraviolet (UV) radiation affects only the surface of molecular clouds (i.e., photo-dissociation regions (PDRs); see, e.g., \citealt{Tielens85a}).
Hence, the primary energy source of molecular clouds with $A_{\mathrm V}>$5~mag is CRs, which indicates the importance of CRs' role to understand the evolution of the ISM in molecular clouds.

Gamma-ray supernova remnants (SNRs) are important targets to study CR chemistry.
Among them, W51C is a middle-aged SNR (3$\times$10$^4$~yrs) at a distance of $d$=5.4~kpc (\citealt{Sato10}), showing bright X-ray and $\gamma$-ray emission (\citealt{Koo95, Abdo09, Aleksic12}).
The $\gamma$-ray emission originates from the interaction between accelerated CRs and molecular clouds (\citealt{Aleksic12}).
Notably, the presence of such interaction was confirmed in radio observations.
\citet{Ceccarelli11} observed DCO$^+$(2--1) at five locations in the vicinity of W51C with the IRAM 30-m telescope and determined the CR ionization rate ($\zeta$) at ``point E'' to be $\zeta\sim$10$^{-15}$~s$^{-1}$, which is two orders of magnitude higher than that in the standard Galactic environment ($\zeta\sim$10$^{-17}$~s$^{-1}$).
They argued that $\zeta$ is enhanced by freshly accelerated low-energy CRs.
To date, a high $\zeta$ inside a molecular cloud associated with an SNR has been identified only in W51C and W28 (\citealt{Vaupre14}).
W51C is, therefore, a good target to study CR chemistry.

However, measurement of $\zeta$ in the vicinity of an SNR is not a trivial task.
The CR probes used in previous studies (e.g., DCO$^+$(2--1) as in \citealt{Ceccarelli11}) are very weak and thus are not suitable for wide-area mapping observations of molecular clouds, which are required for an extensive study of $\zeta$ in the vicinity of an SNR.
\citet{Bisbas17} demonstrated that [C$\;${\sc i}] is a good probe of CRs on the basis of 3-dimensional PDR model calculations.
They found that the C$^0$/CO abundance ratio ($X_{\mathrm{C/CO}}$) would be $\sim$0.1 in the standard Galactic environment, whereas $X_{\mathrm{C/CO}}$ would be enhanced up to $\sim$10 in the extreme environment with $\zeta\sim10^{-15}~\mathrm{s}^{-1}$.
The C$^0$ abundance is enhanced due to the destruction of CO by He$^+$, which is produced by CRs.
It is well known that both [C$\;${\sc i}] and CO show intense emission lines in the (sub-)millimeter wavelength range.
Hence, [C$\;${\sc i}] and CO are suitable for measuring $\zeta$ in the vicinity of W51C.
\citet{Arikawa99} observed [C$\;${\sc i}] ($^3P_1$--$^3P_0$) and $^{12}$CO(3--2) in the W51 complex region with the Mt. Fuji sub-millimeter telescope and found that [C$\;${\sc i}]/$^{12}$CO(3--2) intensity ratio is high ($\sim$0.6) in the vicinity of W51C.
However, the relatively poor spatial grid size of 3$\arcmin$ (3.4~pc) in their observations did not allow them to resolve fully the spatial structures of the molecular clouds.
In order to distinguish the CR-originating [C$\;${\sc i}] enhancement inside molecular clouds from the UV-originating [C$\;${\sc i}] enhancement at the surface, mapping observations in a higher spatial resolution are essential.

In this paper, we present [C$\;${\sc i}] ($^3P_1$--$^3P_0$), $^{12}$CO(1--0), and $^{13}$CO(1--0) maps of W51C.
The mapping area is 1500$\arcsec\times$750$\arcsec$ (28~pc$\times$14~pc), which is one of the most extensive [C$\;${\sc i}] observations of a Galactic SNR.
The spatial resolution is $\sim$18$\arcsec$ (0.3~pc), which is sufficiently high to spatially resolve the structures of the observed molecular clouds.
Analyzing the data, we evaluate the effects of CRs on $X_{\mathrm{C/CO}}$ inside the molecular clouds associated with W51C.

\section{Observations and Data Reduction}

The observations of [C$\;${\sc i}] ($^3P_1$--$^3P_0$) were carried out during the period of 2019 August 12 to 17 with the Atacama Submillimeter Telescope Experiment (ASTE) 10-m telescope (\citealt{Ezawa04}).
The total observation time was 21 hrs, in which 10 hrs were on source.
The mapping area is shown in figure~\ref{obsregion}.
Nine sub-regions with an area of 250$\arcsec\times$250$\arcsec$ were covered with the on-the-fly (OTF) mapping using the ALMA Band 8 receiver (\citealt{Satou08}) and backend WHSF (\citealt{Iguchi08, Okuda08}).
The frequency resolution and bandwidth were 500 kHz and 1024 MHz, respectively.
The typical system temperature including the atmosphere was 1100~K.
The pointing accuracy of the telescope was kept to be $<2\arcsec$, using observations of R Aquila in $^{12}$CO(3--2).

Simultaneous observations of $^{12}$CO(1--0) and $^{13}$CO(1--0) were carried out during the period of 2020 January 21 to March 25 with the Nobeyama 45-m telescope.
The total observation time was 9 hrs, in which 3.6 hrs were on source.
One region with an area of 1500$\arcsec\times$750$\arcsec$ was covered with the OTF mapping using the four-beam receiver FOREST (\citealt{Minamidani16a}) and backend SAM45 (\citealt{Kuno11}).
The frequency resolution and bandwidth were 30.52 kHz and 62.5 MHz, respectively.
The typical system temperatures including atmosphere were 400~K and 150~K for $^{12}$CO(1--0) and $^{13}$CO(1--0), respectively.
The pointing accuracy of the telescope was kept to be $<4\arcsec$, using observations of RT Aquila in SiO(1--0) maser.

Data reduction of [C$\;${\sc i}], $^{12}$CO(1--0), and $^{13}$CO(1--0) was performed with the NOSTAR software.
For each map, we split data into an array and subtracted the baselines, using a first-order polynomial.
The velocity ranges to subtract baselines were 30--50 and 85--90 km~s$^{-1}$ for [C$\;${\sc i}], $-$40--0, 32--35, and 100--110 km~s$^{-1}$ for $^{12}$CO(1--0), and $-$40--0 and 80--110 km~s$^{-1}$ for $^{13}$CO(1--0).
The amplitude of line intensities was calibrated, using the main beam efficiency of $\eta_{\mathrm{mb}}$=0.45 for [C$\;${\sc i}], 0.35 for $^{12}$CO(1--0), and 0.39 for $^{13}$CO(1--0). 
We used a spatial grid of 5$\farcs$5 and a velocity grid of 0.5 km~s$^{-1}$.
A Bessel-Gauss function was used for convolution.
The effective angular resolution was 18$\arcsec$ for [C$\;${\sc i}] and 17$\arcsec$ for $^{12}$CO(1--0) and $^{13}$CO(1--0).
The typical noise levels of the final cubes were 1.0~K, 0.88~K, and 0.31~K for [C$\;${\sc i}], $^{12}$CO(1--0), and $^{13}$CO(1--0), respectively, in the $T_{\mathrm{mb}}$ scale.

\section{Results}

Figure~\ref{sekibun} shows integrated intensity maps of [C$\;${\sc i}], $^{12}$CO(1--0), and $^{13}$CO(1--0), where the data are integrated over
the velocity range $V_{\mathrm{LSR}}=$55--80~km~s$^{-1}$, which is associated with the W51 complex (\citealt{Parsons12}).
The overall spatial distributions, primarily extending in the east-west direction, are similar among the three maps.
The morphological similarity between [C$\;${\sc i}] and CO maps has been reported in previous studies of Galactic star-forming regions (\citealt{Ikeda02, Kamegai03, Shimajiri13, Izumi21}).
A more detailed look, however, reveals some differences in local structures among them.
Notably, the local maximum is located near the center in the [C$\;${\sc i}] map, whereas it is on the western side in the $^{12}$CO(1--0) and $^{13}$CO(1--0) maps.
The differences in the local structures indicate that $X_{\mathrm{C/CO}}$ is not uniform over the mapping area.

$X_{\mathrm{C/CO}}$ is equivalent to the column density ratio of C$^0$ to CO (i.e., $N$(C$^0$)/$N$(CO)).
We estimated $N$(C$^{0}$) and $N$(CO) from the [C$\;${\sc i}], $^{12}$CO(1--0), and $^{13}$CO(1--0) data in $V_{\mathrm{ LSR}}=$55--80~km~s$^{-1}$, where we assumed the local thermodynamic equilibrium and followed the procedures described in \citet{Izumi21}.
$N$(C$^{0}$) and $N$(CO) were estimated using the following equations;
\begin{equation}
N(\mathrm{C^0}) = 4.67 \times 10^{16} \frac{1+3\exp(-23.6/T_\mathrm{{ex}}) +5\exp(-62.5/T_\mathrm{{ex}})}{1-\exp(-23.6/T_\mathrm{{ex}})} \int \tau_\mathrm{[CI]} \; dv,
\end{equation}
\begin{equation}
N(\mathrm{CO}) = [\mathrm{^{12}CO}]/[\mathrm{^{13}CO}] \times 2.42\times10^{14} \frac{T_\mathrm{ex}+0.87}{1-\exp(-5.29/T_\mathrm{ex})} \int \tau_{\mathrm{^{13}CO}} \; dv,
\end{equation}
where $T_{\textrm{ex}}$ is excitation temperature, $\tau_{\mathrm{[CI]}}$ and $\tau_{\mathrm{^{13}CO}}$ are optical depths of [C$\;${\sc i}] and $^{13}$CO(1--0), respectively, and $[\mathrm{^{12}CO}]/[\mathrm{^{13}CO}]$ is the $^{12}$CO/$^{13}$CO abundance ratio of 71 (\citealt{Frerking82}).
We used $T_{\textrm{ex}}$ estimated from the peak brightness temperature of $^{12}$CO(1--0) assuming that $^{12}$CO(1--0) is optically thick; $T_{\textrm{ex}}$ is estimated on a pixel-by-pixel basis and is common to the estimation of $N$(C$^0$) and $N$(CO).
The minimum and maximum values of $T_{\textrm{ex}}$ in the mapping area are 6~K and 81~K, respectively.
Consequently, the optical depths of [C$\;${\sc i}] and $^{13}$CO(1--0), $N$(C$^0$), and $N$(CO) were derived to be 0.14--1.41, 0.09--0.90, (2--22)$\times$10$^{17}$ cm$^{-2}$, and (3--310)$\times$10$^{17}$ cm$^{-2}$, respectively.

We find that the derived $X_{\mathrm{C/CO}}$ varies in a range of $0.02 < X_{\mathrm{C/CO}} < 1.5$ with a median value of 0.06, which is roughly consistent with those in Galactic star-forming regions (DR15: \citealt{Oka01}; $\rho$ Oph: \citealt{Kamegai03}; Orion: \citealt{Ikeda02}; RCW38: \citealt{Izumi21}).
In the following analyses and discussion, considering that $X_{\mathrm{C/CO}}$ is enhanced by the photodissociation of CO at the surface of molecular clouds, we focus on the regions with $A_V>$100~mag, where $A_V$ is calculated as $A_V$=$N$(CO)$\times$10$^4$/9.4$\times$10$^{20}$ (\citealt{Frerking82}), in order to examine solely the effect of CRs independently of that of interstellar UV radiation.
Figure~\ref{ratio}(a) shows the map of $X_{\mathrm{C/CO}}$, where the regions with $A_\mathrm{V}<$100~mag are masked out.
$X_{\mathrm{C/CO}}$ varies in a range of $0.02 < X_{\mathrm{C/CO}} < 0.16$ (0.05 in median).
We find that $X_{\mathrm{C/CO}}$ is systematically high at the center of the mapping area and is low at the edge.
The gradient suggests that $X_{\mathrm{C/CO}}$ is locally enhanced near the W51C center, which is located at the center bottom of the mapping area.
Figure~\ref{ratio}(b) shows the distribution of $X_{\mathrm{C/CO}}$ plotted against the projected distance from the W51C center, which visualizes a clear enhancement of $X_{\mathrm{C/CO}}$ by a factor of four near the W51C center; $X_{\mathrm{C/CO}}$ is nearly constant of 0.04 at the region far from the W51C center ($>15\arcmin$), while $X_{\mathrm{C/CO}}$ is enhanced up to 0.14 near the W51C center.
This type of clear trend of $X_{\mathrm{C/CO}}$ has never been reported in any other SNRs.

\section{Discussion and conclusions}

\subsection{Origin of the local enhancement of $X_{\mathrm{C/CO}}$}

Focusing on the regions with $A_V>100$~mag in the vicinity of W51C, we have found a characteristic feature of $X_{\mathrm{C/CO}}$: local enhancement of $X_{\mathrm{C/CO}}$ by a factor of four near the W51C center.
Our wide-area maps of [C$\;${\sc i}] and CO greatly contribute to the advancement of our understanding of the global variations of $X_{\mathrm{C/CO}}$.
In the following discussion, we discuss the origin of the local enhancement of $X_{\mathrm{C/CO}}$.

In our analysis, we selected the regions with $A_V>$100~mag, where the effects of interstellar UV radiation to $X_{\mathrm{C/CO}}$ are expected to be negligible.
Here, we validate the assumption, i.e., the interstellar UV radiation does not reach deep inside molecular clouds with a high $A_V$.
Figure~\ref{tempcore}(a) shows a map of the dust temperature $T_\mathrm{d}$, which is estimated from the ratio of Herschel 70-$\micron$ and 160-$\micron$ maps under an assumption of the dust emissivity power-law index of $\beta$=2.
We find that $T_\mathrm{d}$ ranges 20--38~K.
Comparison of $T_\mathrm{d}$ and $A_V$ maps indicates that $T_\mathrm{d}$ is low in the high-$A_V$ regions.
In addition, the distribution of $T_\mathrm{d}$ against the projected distance from the W51C center (figure~\ref{tempcore}(b)) does not show local enhancement near the W51C center.
Hence, the molecular clouds with $A_V>$100~mag are well shielded from interstellar UV radiation and its contribution is negligible for the spatial variations of $X_{\mathrm{C/CO}}$ inside molecular clouds.

Generally, $X_{\mathrm{C/CO}}$ in high-$A_V$ molecular clouds is determined by the ratio of the CR ionization rate ($\zeta$) to the H$_2$ density ($n_{\mathrm{H_2}}$) on the basis of the balance between molecular formation and destruction.
Accordingly, the enhancement of $X_{\mathrm{C/CO}}$ is, if present, caused by a low $n_{\mathrm{H_2}}$ or high $\zeta$.
To investigate the former possibility, we examine the spatial variations of $n_{\mathrm{H_2}}$ of cloud cores.
\citet{Parsons12} presented a catalog of cloud cores, compiled from their $^{13}$CO(3--2) observations.
We plot in figure~\ref{tempcore}(c) the positions of the cataloged cloud cores overlaid on the $^{13}$CO(1--0) integrated intensity map; the plot shows that the cloud cores are identified across the entire mapping area.
Using the ratio between the H$_2$ column density and core diameter, we estimate $n_{\mathrm{H_2}}$ averaged in each core, $\langle n_{\mathrm{H_2}}\rangle$.
Figure~\ref{tempcore}(d) shows $\langle n_{\mathrm{H_2}} \rangle$ plotted against the projected distance from the W51C center.
We find that $\langle n_{\mathrm{H_2}} \rangle$ has a typical value of $\sim$10$^3$~cm$^{-3}$ and shows no clear variation against the projected distance.
Hence, the local enhancement of $X_{\mathrm{C/CO}}$ in figure~\ref{ratio}(b) is likely to be caused not by a low $n_{\mathrm{H_2}}$ but by a high $\zeta$.
Note that the region closest to the W51C center locally has a lower $\langle n_{\mathrm{H_2}} \rangle$ than other regions by a factor of two, which may cause a part of the enhancement of $X_{\mathrm{C/CO}}$ at the region.

Past $\gamma$-ray observations (\citealt{Abdo09}) suggested that W51C is an accelerator of CRs.
The hypothesis is consistent with a high $\zeta$, as probed by $X_{\mathrm{C/CO}}$ in this study.
Then, we quantitatively compare the diffusion distance of CRs with the size of the enhanced area of $X_{\mathrm{C/CO}}$.
In figure~\ref{ratio}(b), $X_{\mathrm{C/CO}}$ is enhanced up to a projected distance of 15$\arcmin$, or 17~pc.
The diffusion distance ($d$) of CRs is expressed in the form of $d=\sqrt{4D(E, B)t}$, where $D(E, B)$ is the diffusion coefficient as a function of the CR energy ($E$) and magnetic field strength ($B$) and $t$ is the time after the supernova explosion.
Using $D(E, B)$ in \citet{Gabici09}, $E$=100~MeV, $B=3$~$\mu$G, and $t=3\times10^4$~yrs (\citealt{Koo95}), we estimate the diffusion distance to be  
\begin{equation}
d=20 \, \left(\frac{E}{100~\mathrm{MeV}}\right)^{0.5} \, \left(\frac{B}{3~\mathrm{\mu G}}\right)^{-0.5} \, \left(\frac{t}{3\times10^4~\mathrm{yr}}\right)^{0.5}~\mathrm{pc},
\end{equation}
which is roughly consistent with the observed value of 17~pc.
Hence, we conclude that the low-energy CRs accelerated in W51C spread to the diffusion distance and enhance $X_{\mathrm{C/CO}}$ in the region.

\subsection{Estimation of $\zeta$ with the time-dependent PDR simulation}

To quantitatively understand the impact of $\zeta$ on $X_{\mathrm{C/CO}}$ and estimate $\zeta$ in the vicinity of W51C, we simulate the chemical evolution in the molecular cloud ($l$, $b$)=(+49:09:09, $-$00:22:03), which shows $X_{\mathrm{C/CO}}$=0.14 and $A_V=$170~mag in the cloud center.
We run one-dimensional plane parallel PDR models (\citealt{furuya22}), which solve the time evolution of the gas temperature and abundances of chemical species self-consistently for given gas density distribution and dust-temperature distribution, considering gas-phase and gas-grain interaction (i.e., adsorption and desorption) and heating and cooling processes.
Elemental abundances are taken from \citet{aikawa99}. 
At the initial chemical state, we assume all hydrogen is in H${_2}$, all carbon is in CO, and the remaining oxygen is in H$_2$O ice.
The other elements are assumed to exist as either atoms or atomic ions in the gas phase.

The model simulation consists of two steps.
In the first step, the model simulates the temporal evolution of the gas temperature and chemical abundances for 10$^7$~yrs, assuming the standard $\zeta$ value of $1\times10^{-17}$~s$^{-1}$.
The chosen timescale of 10$^7$~yrs is arbitrary, but it is long enough for the C${^+}$, C$^0$, and CO abundances to reach the steady-state values.
In the second step, the model follows the evolution for $1\times10^5$~yrs with an enhanced $\zeta$ value in a range between $3\times10^{-17}$~s$^{-1}$ and $1\times10^{-14}$~s$^{-1}$.
The first and second steps mimic conditions just before and after the supernova explosion that generates W51C, respectively.

\begin{figure}[ht!]
\begin{center}
\includegraphics[width=0.532\textwidth]{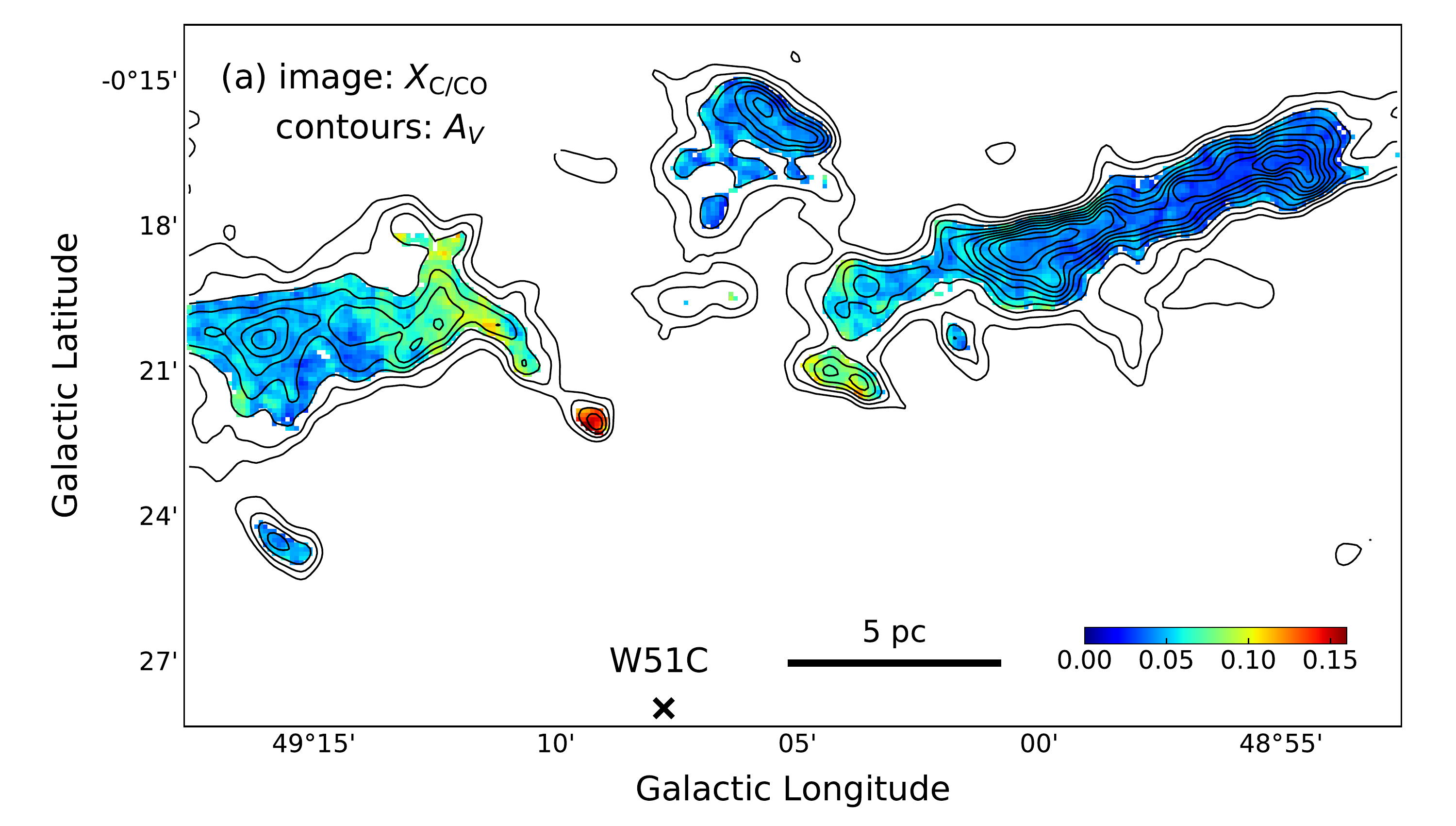}
\includegraphics[width=0.458\textwidth]{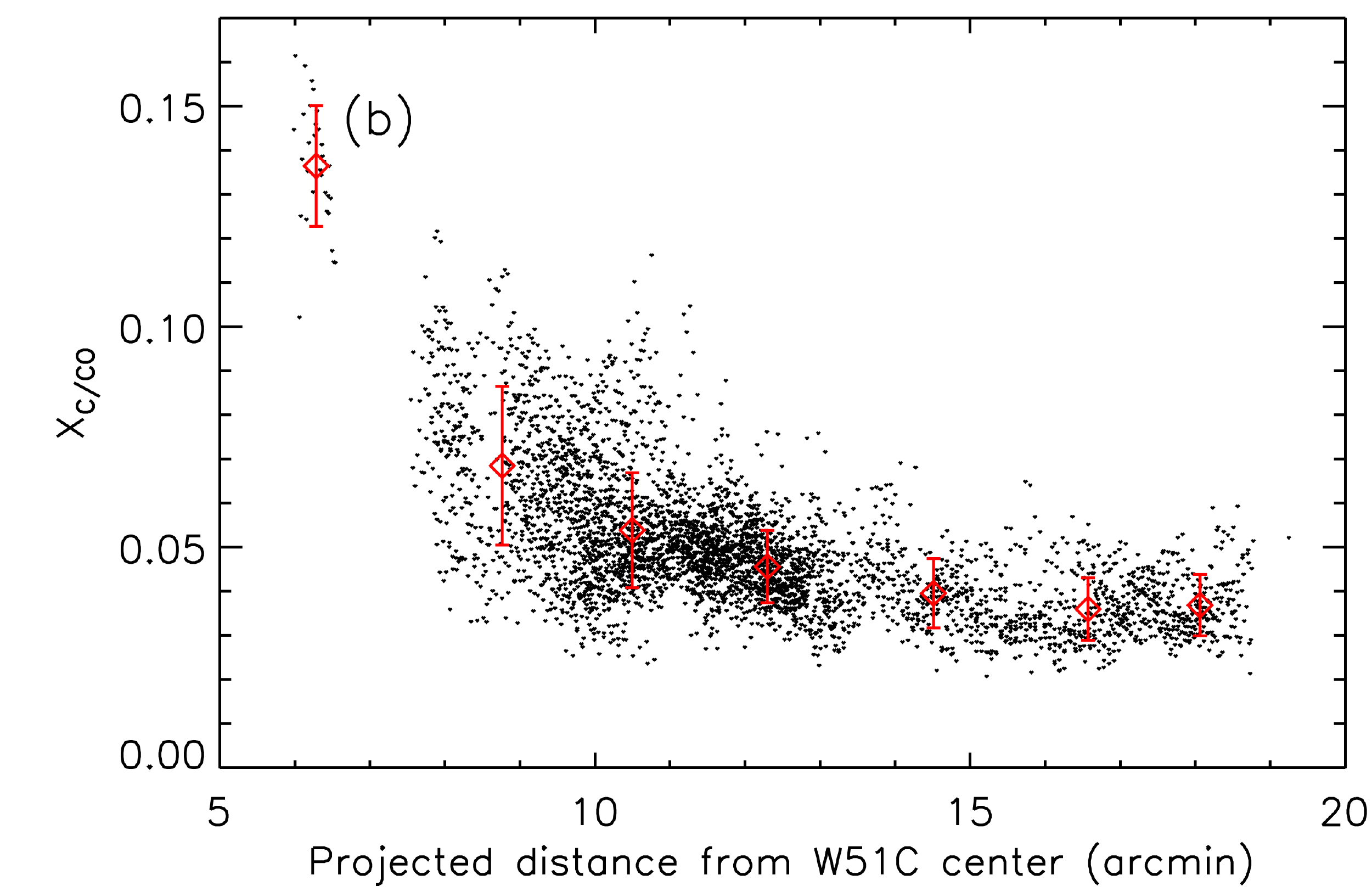}
\end{center}
\caption{(a) Map of the abundance ratio $X_{\mathrm{C/CO}}$. The regions with $A_\mathrm{V}<$100~mag are masked out. Contours indicate $A_V$=50, 75, 100, $\cdots$, 275~mag. The cross near the lower edge indicates the position of the W51C center. (b) $X_{\mathrm{C/CO}}$ plotted against the projected distance from the W51C center. Red data indicate the averages and 1$\sigma$ standard deviations of $X_{\mathrm{C/CO}}$ for 7 sections in 5$\farcm$5--19$\farcm$5. \label{ratio}}
\end{figure}

As input parameters for the model, we employ the UV intensity $G_0$ = 9200, dust temperature $T_\mathrm{d}$=26~K, and the hydrogen (H$\;${\sc i}+H$_2$) gas density $n_{\mathrm{HI+H_2}}$ = 3$\times$10$^3$~cm$^{-3}$.
The UV intensity is estimated using equation (8) in \citet{Hocuk17} with $T_\mathrm{d}$=26~K and $A_V$=85~mag (i.e., half depth of a molecular cloud with $A_V$=170~mag).
The direct effect of the UV radiation is negligible at the cloud center because of the high $A_V$.
$T_\mathrm{d}$ is estimated from figure~\ref{tempcore}(a).
$T_\mathrm{d}$=26~K is consistent with the dust temperature in a molecular cloud achieved by the interstellar radiation field $G_0\sim10^{3-4}$ (\citealt{Tielens05}), where the main heating source of dust in a high-$A_V$ molecular cloud is far-IR radiation from dust in PDR.
$n_{\mathrm{HI+H_2}}$ is determined by a slight optimization of the estimation in figure~\ref{tempcore}(d) to reproduce $X_{\mathrm{C/CO}}$=0.04 at just before the explosion (i.e., observed $X_{\mathrm{C/CO}}$ at the region far from the W51C center, where enhancement of $\zeta$ is likely unaffected. See figure~\ref{ratio}(b).).

\begin{figure}[ht!]
\begin{center}
\includegraphics[width=\textwidth]{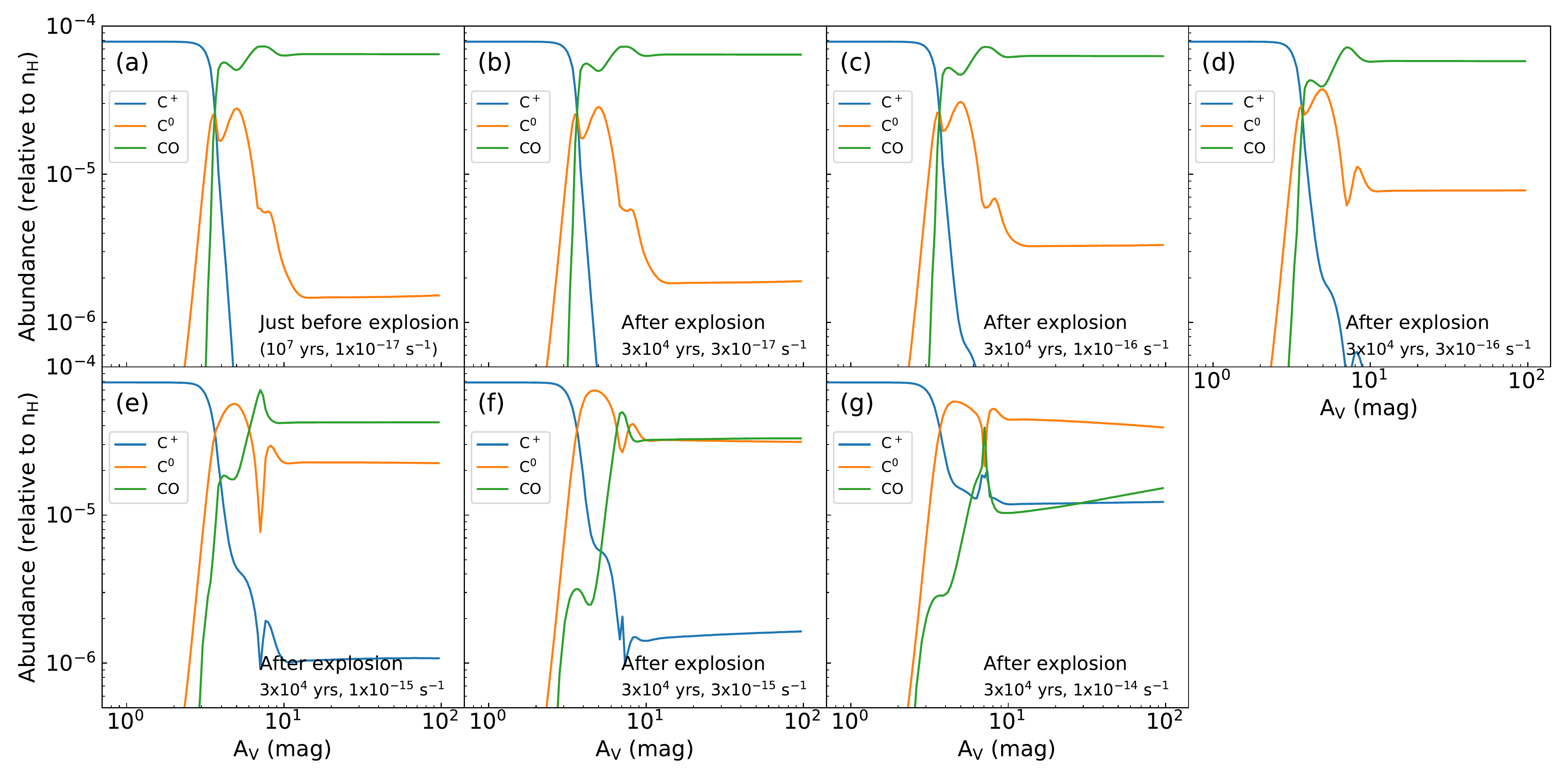}
\end{center}
\caption{Examples of chemical structures predicted by the PDR models. (a) Just before the explosion. (b-g) $3\times10^4$~yrs after the explosion with an enhanced $\zeta$ of $3\times10^{-17}$~s$^{-1}$, $1\times10^{-16}$~s$^{-1}$, $3\times10^{-16}$~s$^{-1}$, $1\times10^{-15}$~s$^{-1}$, $3\times10^{-15}$~s$^{-1}$, and $1\times10^{-14}$~s$^{-1}$.
Blue, orange, and green lines represent the abundances of C$^+$, C$^0$, and CO, respectively.}
\label{pdr1}
\end{figure}

Figure~\ref{pdr1} shows examples of the obtained chemical structures of the cloud just before the supernova explosion and $3\times10^4$~yrs after the explosion assuming an enhanced $\zeta$.
The main difference in the chemical structures is the abundance of C$^0$ (and that of C$^+$) at $A_V>10$~mag, where the interstellar UV radiation field is significantly attenuated.
Higher $\zeta$ causes a higher abundance of C$^0$ in a cloud.
The enhancement of the C$^0$ abundance is due to the destruction of CO by He$^+$, which in turn produces C$^+$ and O (\citealt{Bisbas15}).
Hence, our simulations demonstrate that a non-negligible amount of C$^0$ can be produced from CO and that $X_{\mathrm{C/CO}}$ is enhanced within $3\times10^4$~yrs if $\zeta$ is significantly enhanced after the explosion.
The destruction timescale of CO by He$^+$ ($\tau_{\rm CRchem}$) is described as follows (e.g., \citealt{furuya14});
\begin{equation}
\tau_{\rm CRchem} \sim 2 \times 10^{4} \, \left(\frac{x_{\rm CO}}{1\times 10^{-4}}\right)
\left(\frac{\zeta}{1 \times 10^{-15}~{\rm s^{-1}}}\right)^{-1}~{\rm yrs},
\label{crchem}
\end{equation}
where $x_\mathrm{CO}$ is the abundance of CO relative to H.
This equation supports that an enhanced $\zeta$ realizes the significant destruction of CO at the age of W51C.

\begin{figure}[ht!]
\begin{center}
\includegraphics[width=\textwidth]{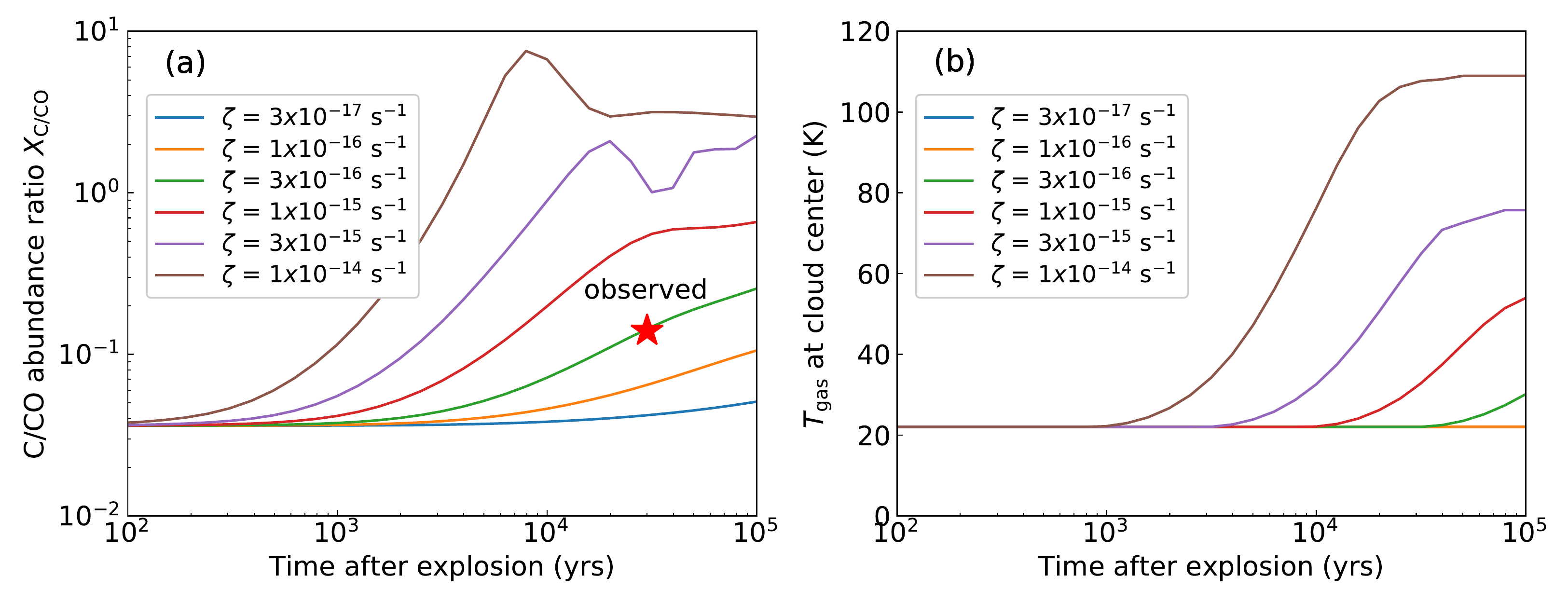}
\end{center}
\caption{Time evolution of (a) the C$^0$/CO abundance ratio $X_{\mathrm{C/CO}}$ and (b) gas temperature $T_g$ at the cloud center after the supernova event that generates W51C. Colors indicates the assumed $\zeta$ values from 3$\times$10$^{-17}$~s$^{-1}$ to 1$\times$10$^{-14}$~s$^{-1}$. The red star indicates the observed $X_{\mathrm{C/CO}}$ (0.14) and the age of W51C (3$\times$10$^4$~yrs).}
\label{pdr2}
\end{figure}

Figure~\ref{pdr2} summarizes the obtained time-evolution curves of $X_{\mathrm{C/CO}}$ and $T_\mathrm{g}$ at the cloud center for a set of $\zeta$, where $X_{\mathrm{C/CO}}$ is evaluated from the integrated abundance of C$^0$ and CO from $A_V=0$~mag to 85~mag to directly compare the model predictions with observed values.
Figure~\ref{pdr2}(a) shows that $X_{\mathrm{C/CO}}$ starts to increase 10$^{2-4}$~yrs after the explosion and reaches a steady state value in $>10^4$~yrs.
Longer time is necessary for lower $\zeta$ to reach the steady state, which is consistent with equation~(\ref{crchem}).
Comparing the observed $X_{\mathrm{C/CO}}$ with our model predictions, we estimate $\zeta$ to be 3$\times$10$^{-16}$~s$^{-1}$.
It is notable that $X_{\mathrm{C/CO}}$ is not in equilibrium and $T_{\mathrm{g}}$ is not enhanced yet at 3$\times$10$^4$~yrs in the condition of $\zeta=3\times10^{-16}$~s$^{-1}$.
These results indicate the importance of time-dependent simulations to examine CR chemistry in SNRs.
We conclude that $\zeta$ is enhanced by a factor of 30 (i.e, from $\sim$10$^{-17}$~s$^{-1}$ to 3$\times$10$^{-16}$~s$^{-1}$) by low-energy CRs accelerated in W51C.

\begin{figure}[ht!]
\begin{center}
\includegraphics[width=0.8\textwidth]{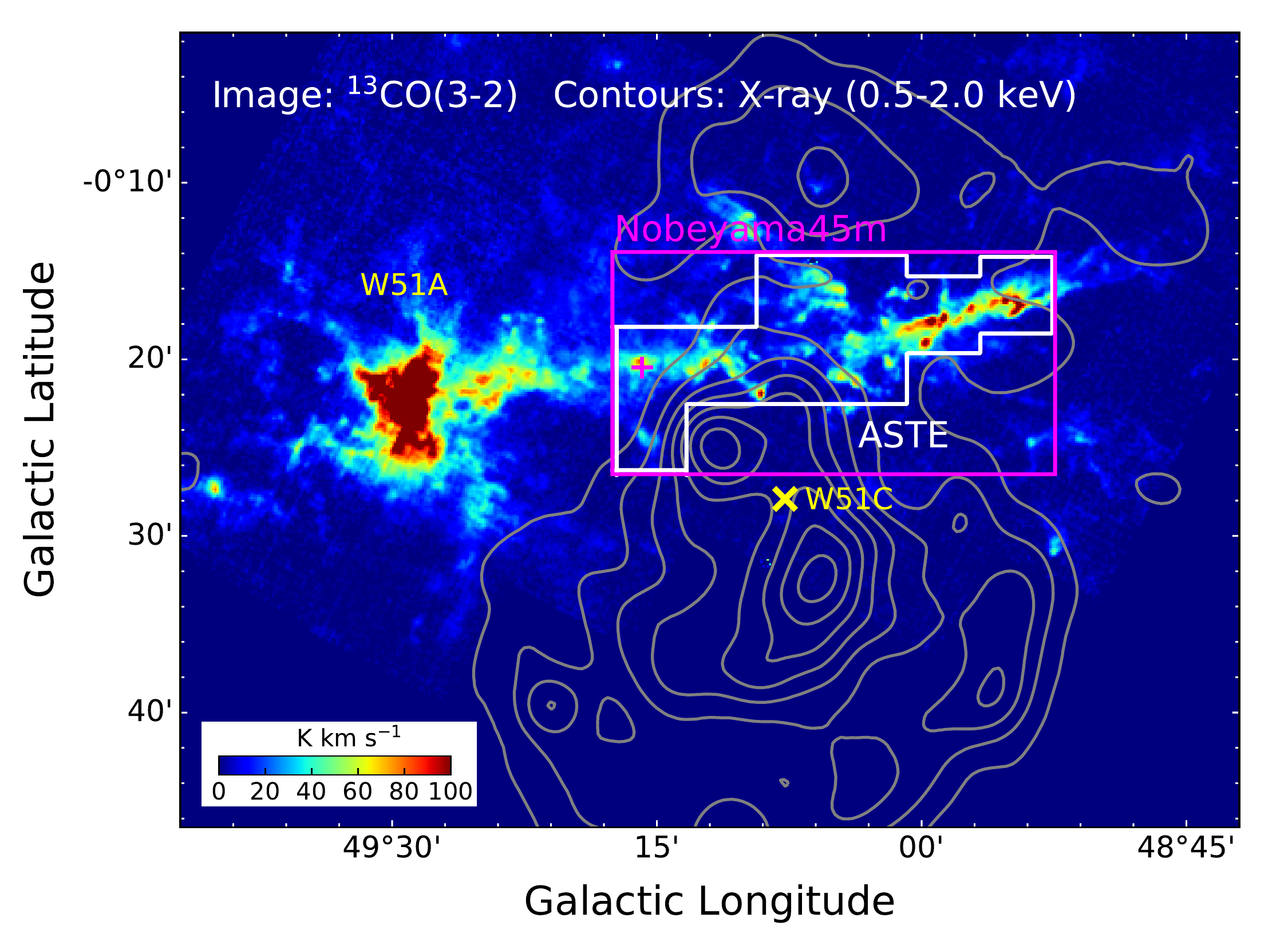}
\end{center}
\caption{Mapping areas with the ASTE and Nobeyama 45-m telescopes overlaid on the integrated intensity map of $^{13}$CO(3--2)  for $V_\mathrm{LSR}$ = 55--80 km~s$^{-1}$ (\citealt{Parsons12}). Contours indicate the ROSAT X-ray image in $E$=0.5--2.0~keV (\citealt{Koo95}) for 8 levels linearly spaced for (2--12)$\times$10$^{-5}$~counts~s$^{-1}$~pixel$^{-1}$. The yellow cross indicates the position of W51C (\citealt{Massaro15}). The magenta plus indicates the position of ``point E" observed by \citet{Ceccarelli11}.
\label{obsregion}}
\end{figure}

This is the first and independent confirmation of the enhancement of $\zeta$, which was first reported by \citet{Ceccarelli11} using DCO$^+$\footnote{\citet{Ceccarelli11} estimated $\zeta$ to be $\sim$10$^{-15}$~s$^{-1}$ at ``point E'' (see figure~\ref{obsregion}). In our result (figure~\ref{ratio}(a)), $X_{\mathrm{C/CO}}$ is $\sim$0.6 at ``point E'', which corresponds to $\zeta=1\times10^{-16}$~s$^{-1}$. The difference of one order of magnitude may be due to the difference in the probes.}.
Furthermore, the enhanced $X_{\mathrm{C/CO}}$ near the W51C center (figure~\ref{ratio}(a)) indicates that $\zeta$ is enhanced not only at the single point reported in \citet{Ceccarelli11} but also in multiple molecular clouds in the vicinity of W51C.
These results demonstrate that [C$\;${\sc i}] is a powerful tool to investigate the interaction between CRs and the ISM for a wide area in the vicinity of SNRs.
Recently, \citet{Tanaka21} found an enhancement of $X_{\mathrm{C/CO}}$ in a surveyed region in the Galactic center, which may be associated with the SNR Sgr A East.
The observed $X_{\mathrm{C/CO}}$ (0.66) is two times higher than the average value in the central molecular zone (0.32).
They estimated $\zeta$ to be $\sim$10$^{-16}$~s$^{-1}$ from C$^0$/CO and CN/HCN abundance ratios.
The present study and \citet{Tanaka21} have advanced the understanding of the interaction between CRs and the ISM in the vicinity of SNRs.
However, the number of detailed studies of CRs in the vicinity of SNRs is still limited.
Further observations of [C$\;${\sc i}] in other SNRs with single-dish sub-mm telescopes or ALMA are essential for a thorough understanding of the ISM evolution in conjunction with CRs in molecular clouds.

\clearpage

\begin{ack}
We express many thanks to the anonymous referee for useful comments.
This work is based on observations with the ASTE and Nobeyama 45-m telescopes.
The ASTE telescope is operated by the National Astronomical Observatory of Japan (NAOJ).
The Nobeyama 45-m telescope is operated by the Nobeyama Radio Observatory, a branch of the NAOJ.
This work is partially based on archival data obtained with the Herschel Space Observatory, which is an ESA space observatory with science instruments provided by the European-led Principal Investigator consortia and with significant participation of NASA.
This work is supported by JSPS KAKENHI Grant No. 19H05075, 21H01136 (HS), 20H05847, and 21H04487 (KF), and NAOJ ALMA Scientific Research Grant No. 2017-06B (MY).
\end{ack}

\begin{figure}[ht!]
\begin{center}
\includegraphics[width=0.7\textwidth]{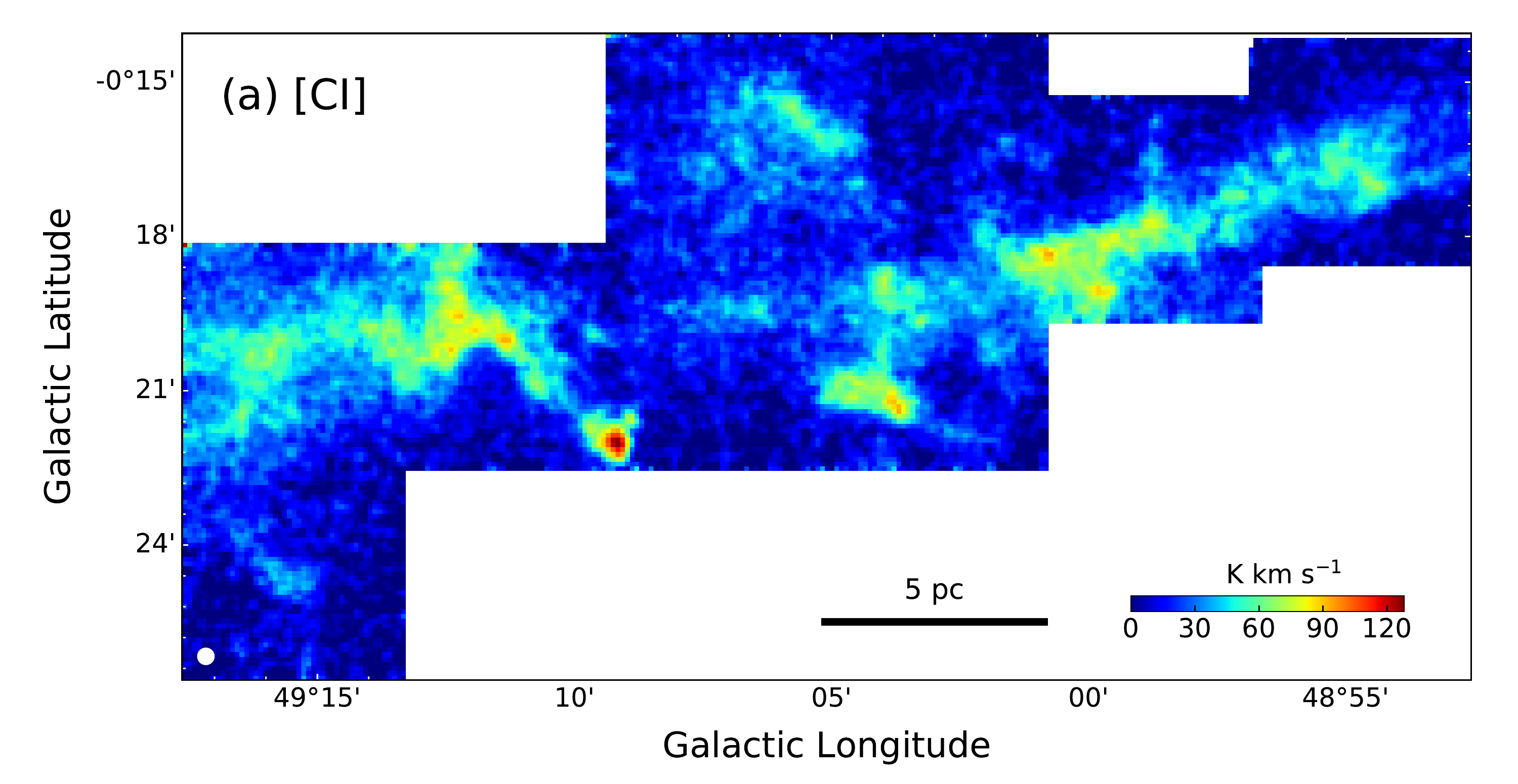}\\
\includegraphics[width=0.7\textwidth]{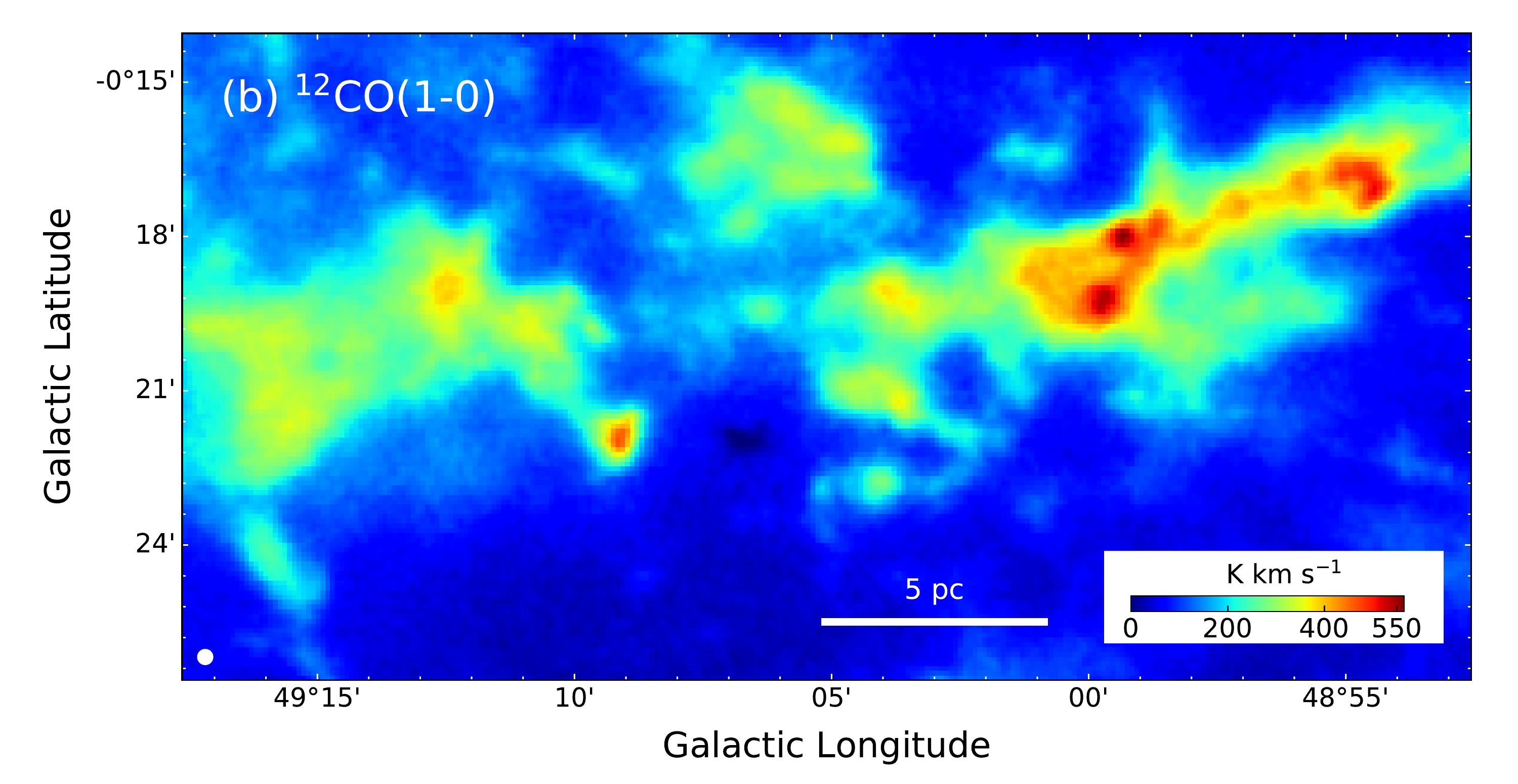}\\
\includegraphics[width=0.7\textwidth]{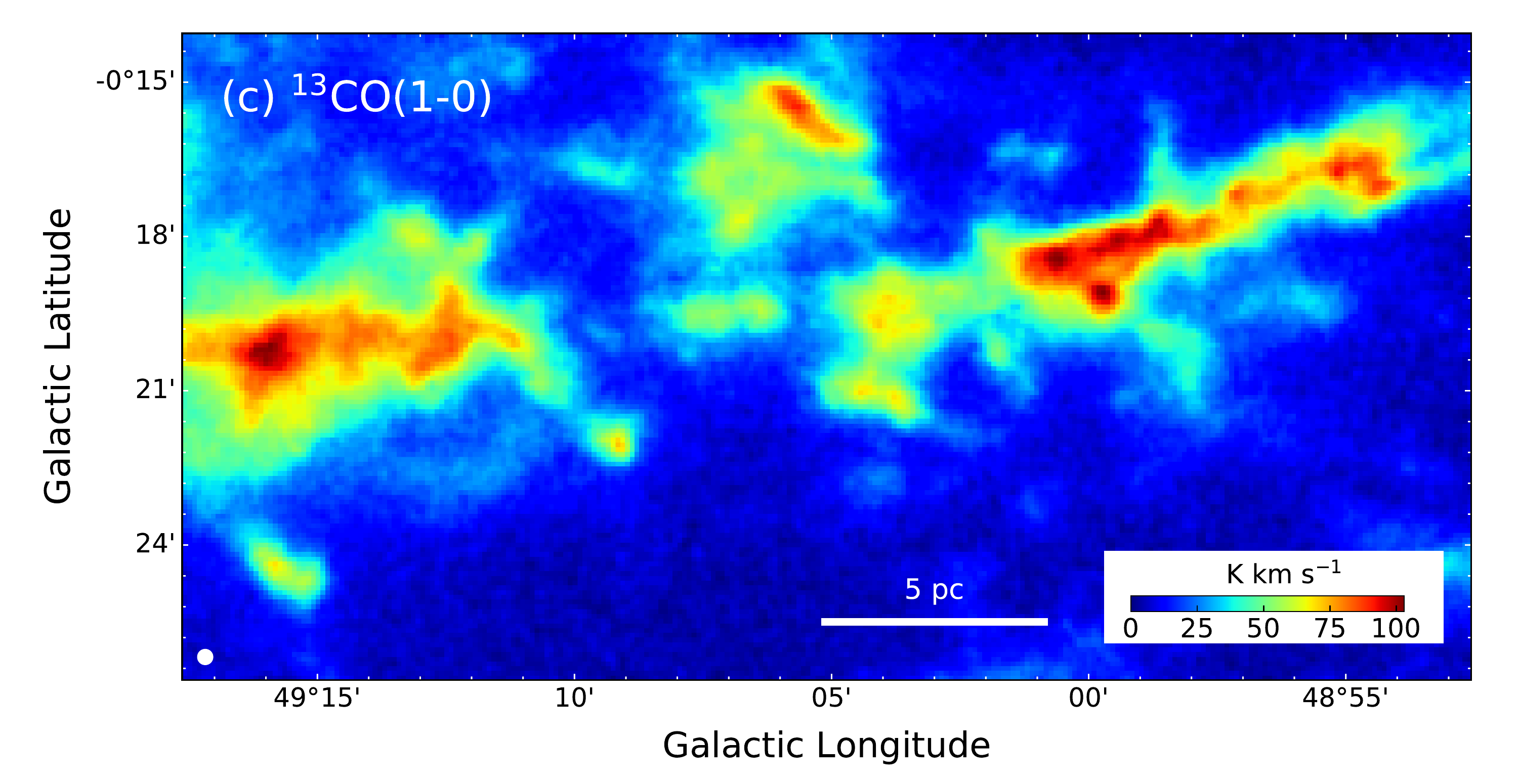}
\end{center}
\caption{Integrated intensity maps of (a) [C$\;${\sc i}], (b) $^{12}$CO(1--0), and (c) $^{13}$CO(1--0) for $V_\mathrm{LSR}$ = 55--80 km~s$^{-1}$. The beam size is indicated at the bottom left corner in each panel. \label{sekibun}}
\end{figure}

\begin{figure}[ht!]
\begin{center}
\includegraphics[width=0.555\textwidth]{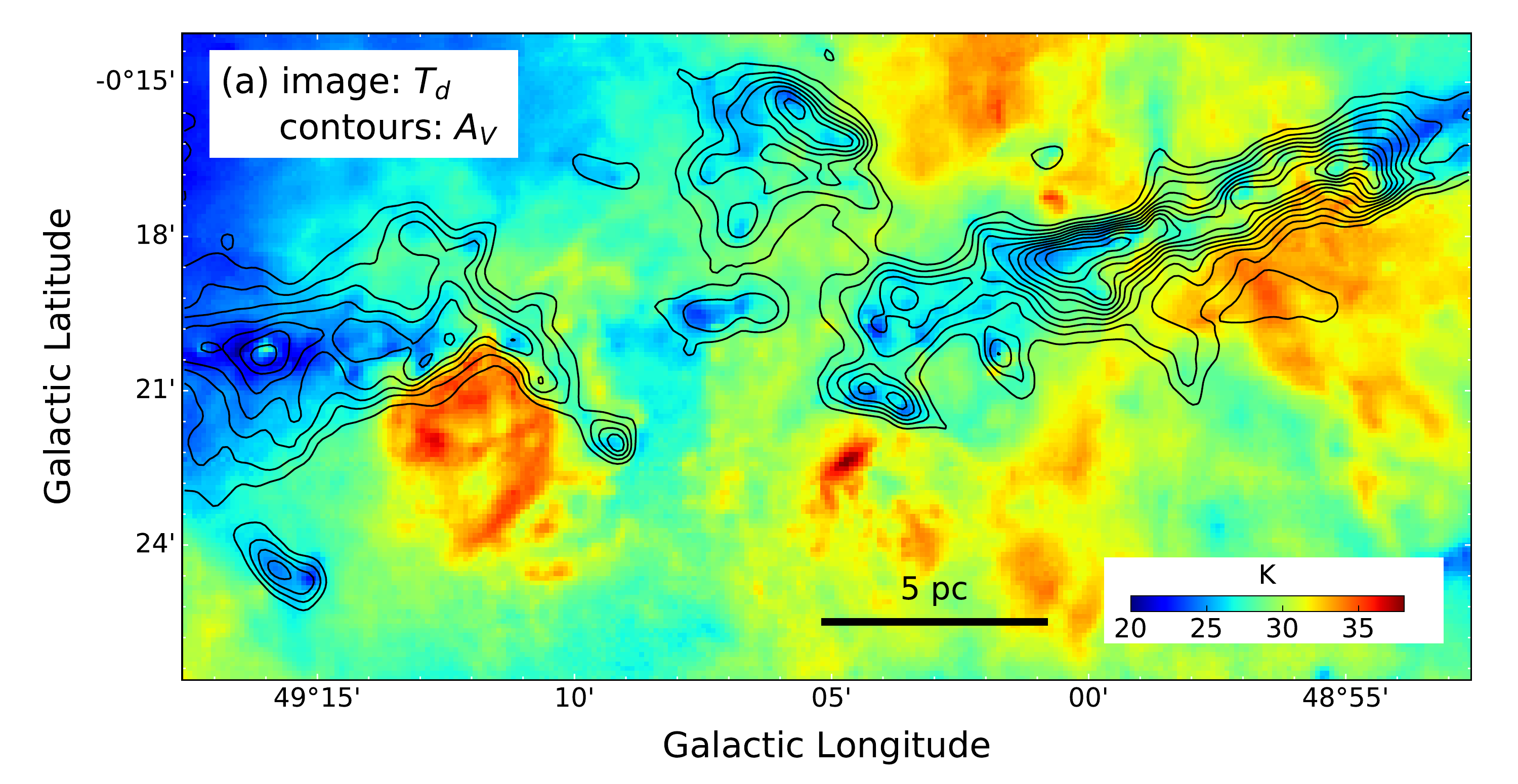}
\includegraphics[width=0.435\textwidth]{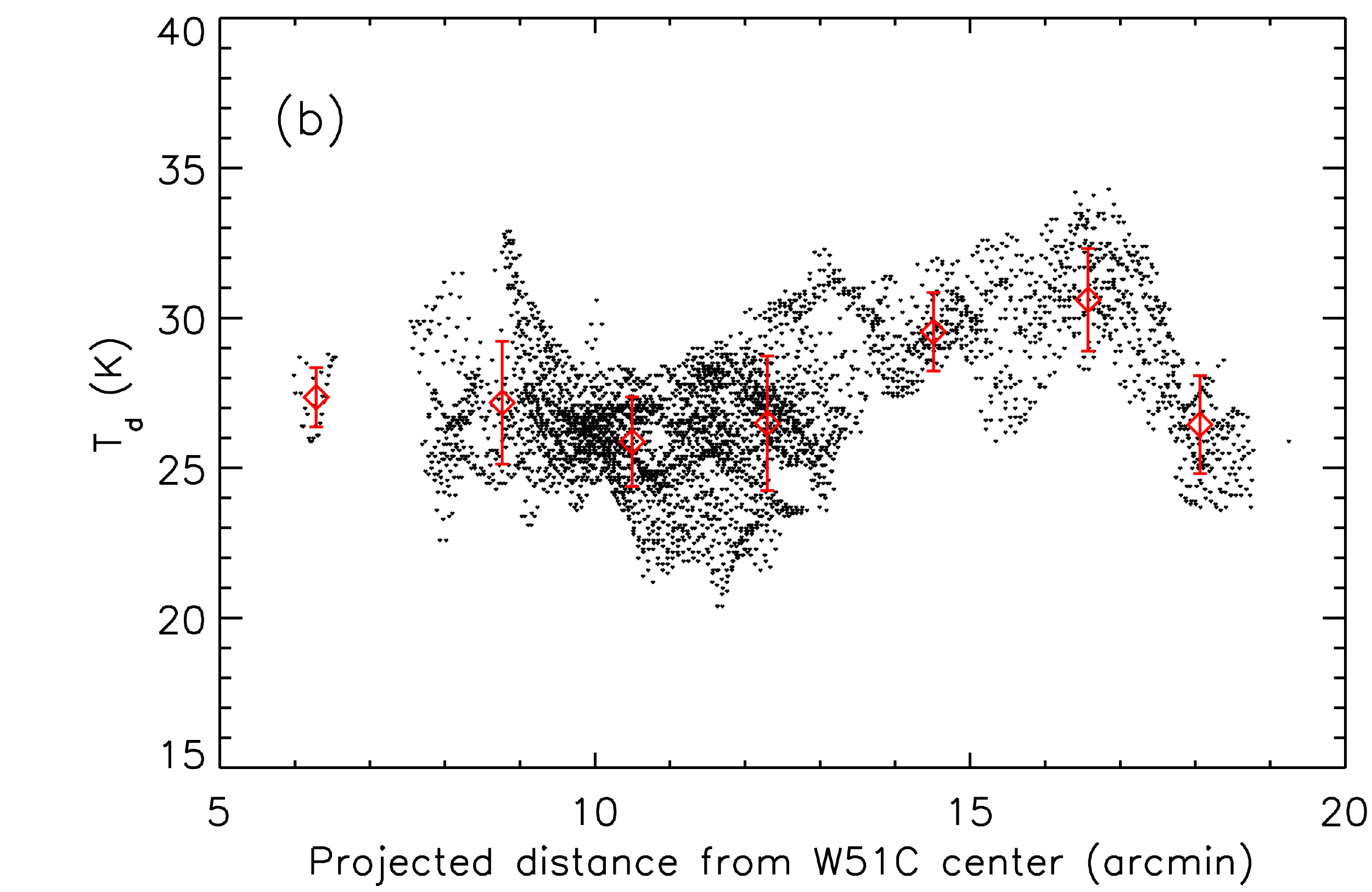}\\
\includegraphics[width=0.555\textwidth]{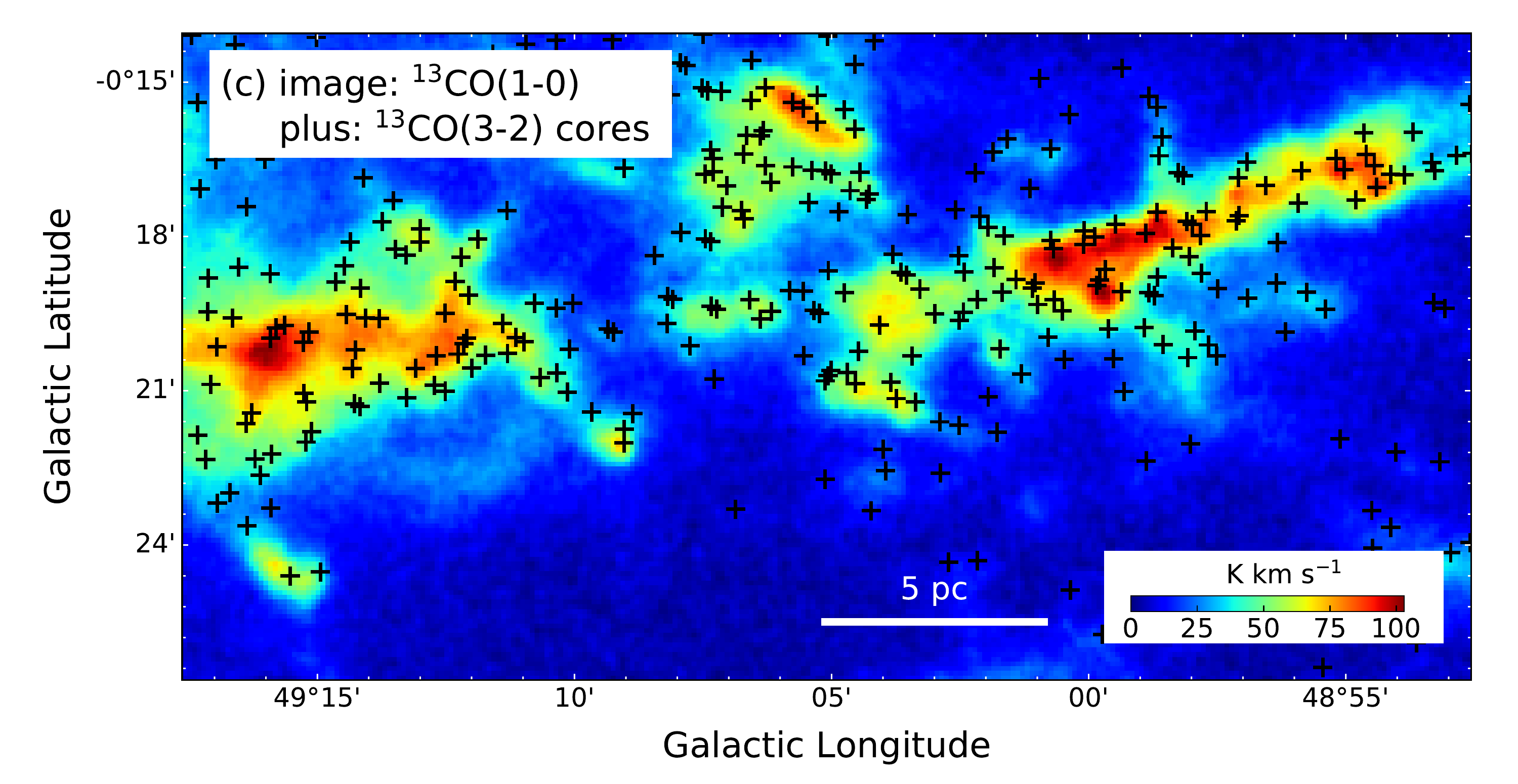}
\includegraphics[width=0.435\textwidth]{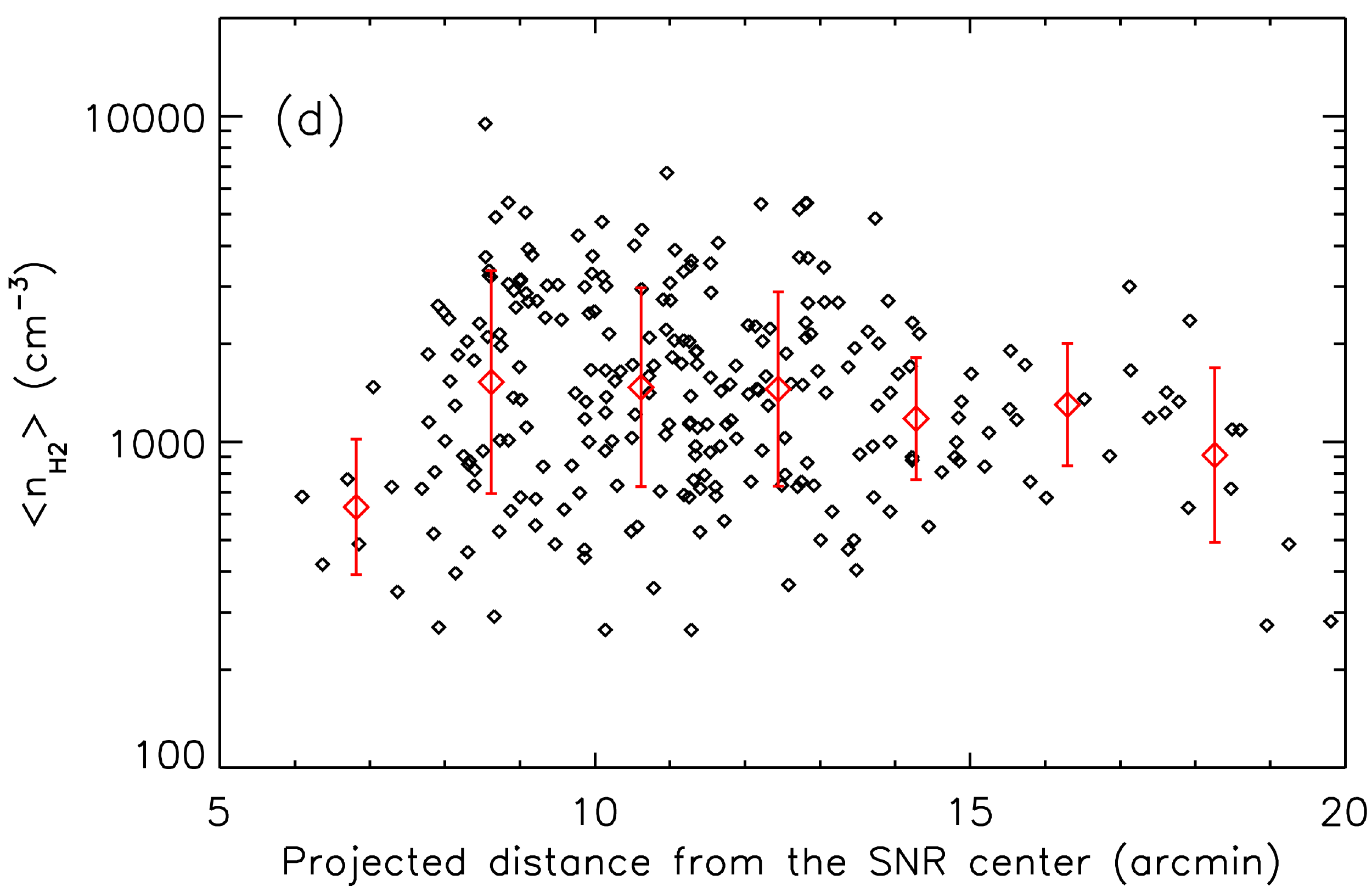}
\end{center}
\caption{(a) Map of the dust temperature $T_\mathrm{d}$. Contours indicate $A_V$ as in figure~\ref{ratio}(a). (b) $T_\mathrm{d}$ plotted against the projected distance from the W51C center, where the data points of the regions with $A_V<100$~mag are excluded. Red data indicate average and 1$\sigma$ standard deviation of the plotted data. (c) Positions of the cloud cores identified in \citet{Parsons12}, overlaid on the $^{13}$CO(1--0) map as in figure~\ref{sekibun}(c). (d) $\langle n_{\mathrm{H_2}} \rangle$ plotted against the projected distance from the W51C center where the data from the regions that are not covered by the ASTE mapping are excluded. Red data indicate the average and 1$\sigma$ standard deviation of the plotted data. \label{tempcore}}
\end{figure}

\clearpage
\appendix
\section{Spectra}

\begin{figure}[ht!]
\begin{center}
\includegraphics[width=\textwidth]{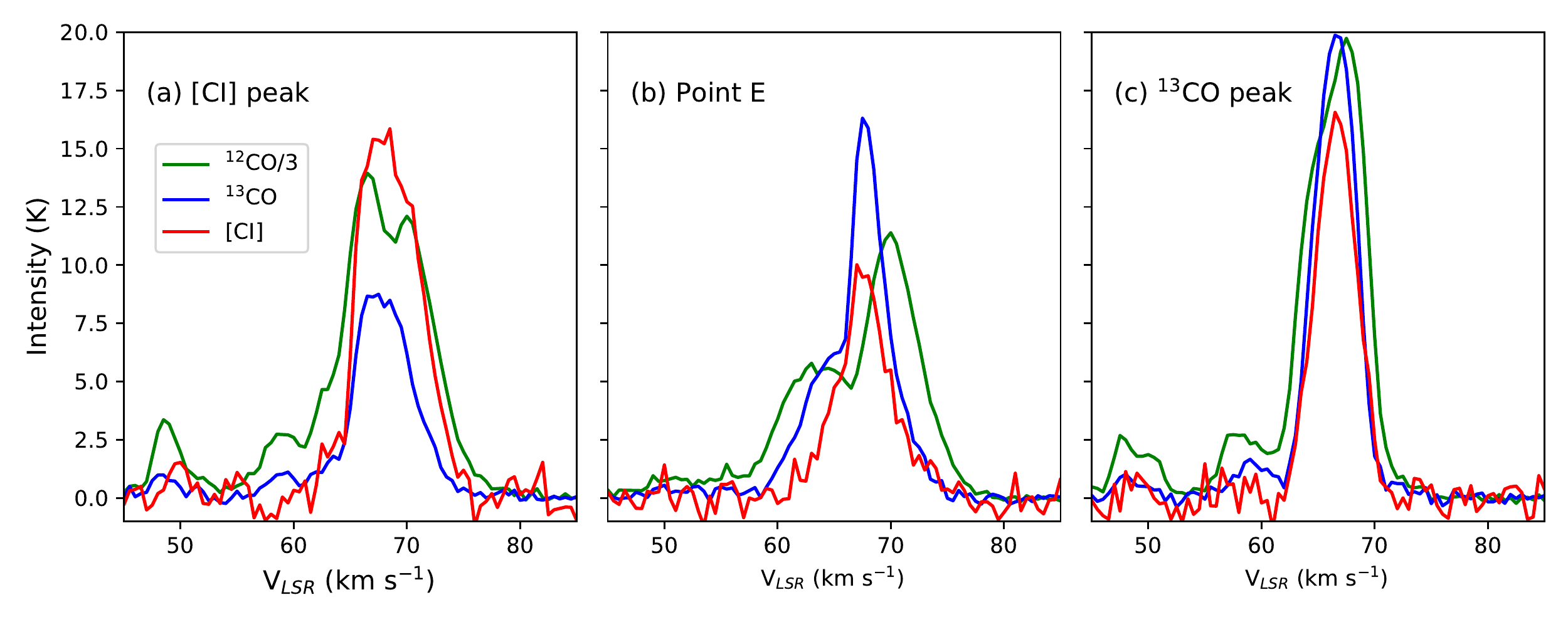}
\end{center}
\caption{[C$\;${\sc i}], $^{12}$CO(1--0), and $^{13}$CO(1--0) spectra for three representative positions. (a) [C$\;${\sc i}] peak in figure~\ref{sekibun}(a); ($l$, $b$)= (+49:09:09.9, -0:22:03.2). (b) ``point E'' in \citet{Ceccarelli11}; ($l$, $b$)= (+49:15:57.4, -0:20:35.0). (c) $^{13}$CO(1--0) peak in figure~\ref{sekibun}(c); ($l$, $b$)= (+49:00:33.6, -0:18:26.6). Spectra are extracted from a 5~pix$\times$5~pix (28$\farcs$5$\times$28$\farcs$5) area centered at the three positions. $^{12}$CO(1--0) spectra are scaled by 1/3 for display purpose.}
\label{spectra}
\end{figure}

Figure~\ref{spectra} shows [C$\;${\sc i}], $^{12}$CO, and $^{13}$CO spectra extracted from three representative positions; the [C$\;${\sc i}] peak in figure~\ref{sekibun}(a), ``point E'' in \citet{Ceccarelli11}, and the $^{13}$CO(1--0) peak in figure~\ref{sekibun}(c).
Velocity structures of [C$\;${\sc i}] are similar to those of $^{13}$CO, but different from those of $^{12}$CO especially in figures~\ref{spectra}(a) and \ref{spectra}(b).
This difference may be caused by the self-absorption of $^{12}$CO(1--0).
Since $N$(C$^0$), $N$(CO), and $X_{\mathrm{C/CO}}$ are not sensitive to $T_{\mathrm ex}$ (\citealt{Izumi21}), underestimation of $T_{\mathrm ex}$ due to the self-absorption of $^{12}$CO(1--0) does not affect our discussion.

\clearpage
\section{$X_{\mathrm{C/CO}}$ vs. $A_V$}

\begin{figure}[ht!]
\begin{center}
\includegraphics[width=0.8\textwidth]{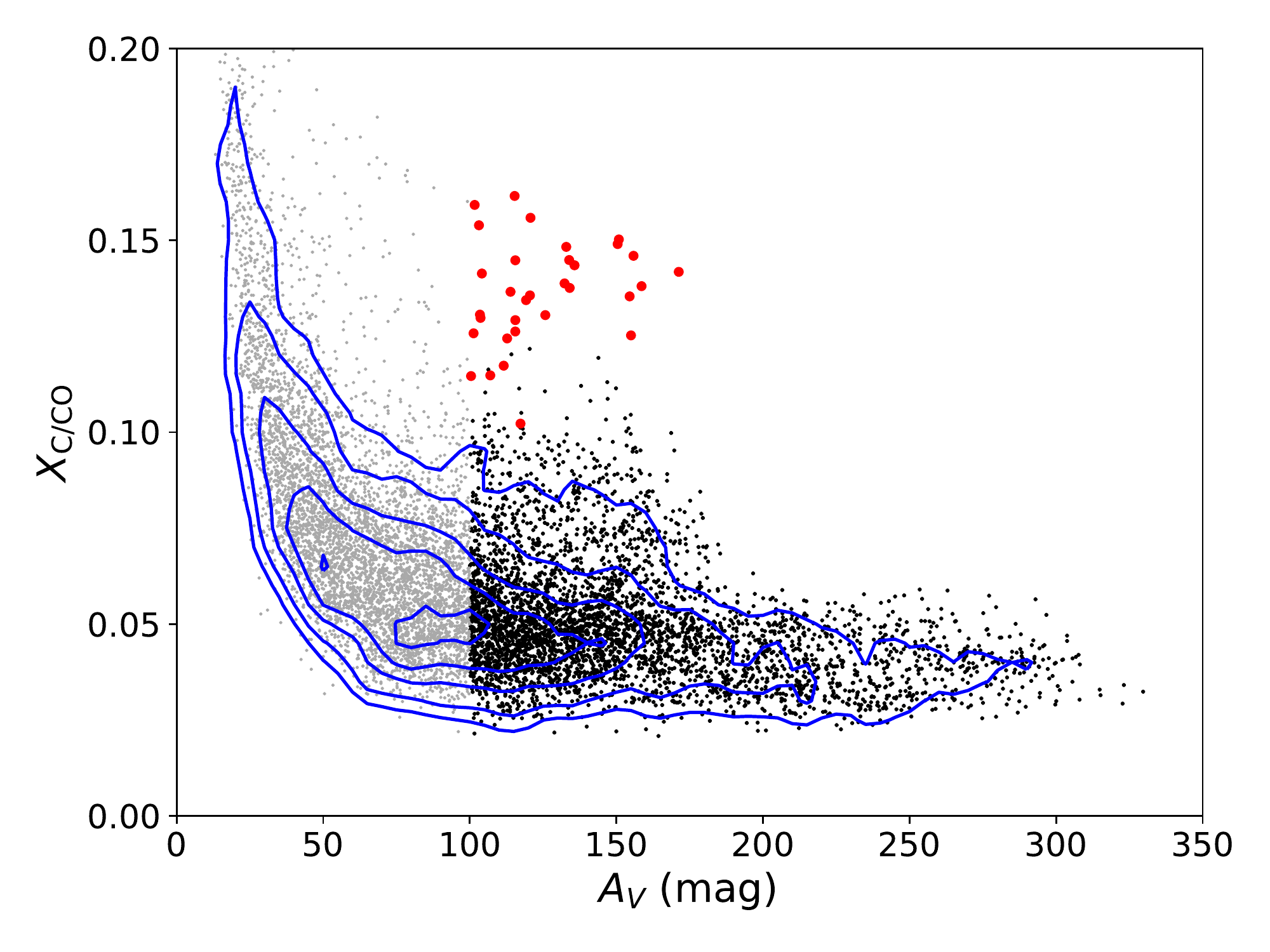}
\end{center}
\caption{$X_{\mathrm{C/CO}}$ ratio plotted against $A_V$. Regions with $A_V<$100~mag are also plotted in gray color as a reference. Blue contours indicate the number density of data points (5, 10, 20, 30, and 40 data points per bin), where the bin size is 5~mag along the x-axis and 0.005 along the y-axis. The data points in red color indicate the molecular cloud closest to the W51C center, which has the maximum $X_{\mathrm{C/CO}}$ (see figure~\ref{ratio}).}
\label{Av_ratio}
\end{figure}

Figure~\ref{Av_ratio} shows $X_{\mathrm{C/CO}}$ plotted against $A_V$.
The overall trend visualized in the blue contours indicates that higher-$A_V$ regions have lower $X_{\mathrm{C/CO}}$ as already reported in \citet{Izumi21}.
The decrement of $X_{\mathrm{C/CO}}$ is rapid in $A_V<$100~mag, while the decrement is negligible in $A_V>$100~mag.
This result supports that the interstellar UV radiation is shielded in $A_V>$100~mag.
Note that the large variation of $X_{\mathrm{C/CO}}$ in $A_V=$100--180~mag is due to the local enhancement of $X_{\mathrm{C/CO}}$ near the W51C center.
We confirm that the molecular cloud closest to the W51C center, which is shown in red color, has higher $X_{\mathrm{C/CO}}$ than other regions with similar $A_V$.

\clearpage
\bibliography{reference.bib}

\begin{thebibliography}{}
\expandafter\ifx\csname natexlab\endcsname\relax\def\natexlab#1{#1}\fi
\providecommand{\url}[1]{\href{#1}{#1}}
\providecommand{\dodoi}[1]{doi:~\href{http://doi.org/#1}{\nolinkurl{#1}}}
\providecommand{\doeprint}[1]{\href{http://ascl.net/#1}{\nolinkurl{http://ascl.net/#1}}}
\providecommand{\doarXiv}[1]{\href{https://arxiv.org/abs/#1}{\nolinkurl{https://arxiv.org/abs/#1}}}

\bibitem[{{Abdo} {et~al.}(2009){Abdo}, {Ackermann}, {Ajello}, {Baldini},
  {Ballet}, {Barbiellini}, {Baring}, {Bastieri}, {Baughman}, {Bechtol},
  {Bellazzini}, {Berenji}, {Blandford}, {Bloom}, {Bonamente}, {Borgland},
  {Bouvier}, {Bregeon}, {Brez}, {Brigida}, {Bruel}, {Burnett}, {Buson},
  {Caliandro}, {Cameron}, {Caraveo}, {Casandjian}, {Cecchi}, {{\c{C}}elik},
  {Chekhtman}, {Cheung}, {Chiang}, {Ciprini}, {Claus}, {Cohen-Tanugi},
  {Cominsky}, {Conrad}, {Cutini}, {Dermer}, {de Angelis}, {de Palma}, {Digel},
  {Dormody}, {Silva}, {Drell}, {Dubois}, {Dumora}, {Farnier}, {Favuzzi},
  {Fegan}, {Focke}, {Fortin}, {Frailis}, {Fukazawa}, {Funk}, {Fusco},
  {Gargano}, {Gasparrini}, {Gehrels}, {Germani}, {Giavitto}, {Giebels},
  {Giglietto}, {Giordano}, {Glanzman}, {Godfrey}, {Grenier}, {Grondin},
  {Grove}, {Guillemot}, {Guiriec}, {Hanabata}, {Harding}, {Hayashida}, {Hays},
  {Hughes}, {Jackson}, {J{\'o}hannesson}, {Johnson}, {Johnson}, {Johnson},
  {Kamae}, {Katagiri}, {Kataoka}, {Katsuta}, {Kawai}, {Kerr}, {Kn{\"o}dlseder},
  {Kocian}, {Kuss}, {Lande}, {Latronico}, {Lemoine-Goumard}, {Longo},
  {Loparco}, {Lott}, {Lovellette}, {Lubrano}, {Makeev}, {Mazziotta}, {McEnery},
  {Meurer}, {Michelson}, {Mitthumsiri}, {Mizuno}, {Moiseev}, {Monte},
  {Monzani}, {Morselli}, {Moskalenko}, {Murgia}, {Nakamori}, {Nolan}, {Norris},
  {Nuss}, {Ohsugi}, {Okumura}, {Omodei}, {Orlando}, {Ormes}, {Paneque},
  {Parent}, {Pelassa}, {Pepe}, {Pesce-Rollins}, {Piron}, {Porter}, {Rain{\`o}},
  {Rando}, {Razzano}, {Reimer}, {Reimer}, {Reposeur}, {Ritz}, {Rodriguez},
  {Romani}, {Roth}, {Ryde}, {Sadrozinski}, {Sanchez}, {Sander}, {Saz
  Parkinson}, {Scargle}, {Schalk}, {Sgr{\`o}}, {Siskind}, {Smith}, {Smith},
  {Spandre}, {Spinelli}, {Strickman}, {Suson}, {Tajima}, {Takahashi},
  {Takahashi}, {Tanaka}, {Thayer}, {Thayer}, {Thompson}, {Tibaldo}, {Tibolla},
  {Torres}, {Tosti}, {Tramacere}, {Uchiyama}, {Usher}, {Vasileiou}, {Venter},
  {Vilchez}, {Vitale}, {Waite}, {Wang}, {Winer}, {Wood}, {Yamazaki}, {Ylinen},
  \& {Ziegler}}]{Abdo09}
{Abdo}, A.~A., {Ackermann}, M., {Ajello}, M., {et~al.} 2009, \apjl, 706, L1

\bibitem[{{Aikawa} \& {Herbst}(1999)}]{aikawa99}
{Aikawa}, Y., \& {Herbst}, E. 1999, \apj, 526, 314

\bibitem[{{Aleksi{\'c}} {et~al.}(2012){Aleksi{\'c}}, {Alvarez}, {Antonelli},
  {Antoranz}, {Asensio}, {Backes}, {Barres de Almeida}, {Barrio}, {Bastieri},
  {Becerra Gonz{\'a}lez}, {Bednarek}, {Berger}, {Bernardini}, {Biland},
  {Blanch}, {Bock}, {Boller}, {Bonnoli}, {Borla Tridon}, {Bretz},
  {Ca{\~n}ellas}, {Carmona}, {Carosi}, {Colin}, {Colombo}, {Contreras},
  {Cortina}, {Cossio}, {Covino}, {Da Vela}, {Dazzi}, {De Angelis}, {De Caneva},
  {De Cea del Pozo}, {De Lotto}, {Delgado Mendez}, {Diago Ortega}, {Doert},
  {Dom{\'\i}nguez}, {Dominis Prester}, {Dorner}, {Doro}, {Eisenacher},
  {Elsaesser}, {Ferenc}, {Fonseca}, {Font}, {Fruck}, {Garc{\'\i}a L{\'o}pez},
  {Garczarczyk}, {Garrido}, {Giavitto}, {Godinovi{\'c}}, {Gonz{\'a}lez
  Mu{\~n}oz}, {Gozzini}, {Hadasch}, {H{\"a}fner}, {Herrero}, {Hildebrand},
  {Hose}, {Hrupec}, {Huber}, {Jankowski}, {Jogler}, {Kadenius}, {Kellermann},
  {Klepser}, {Kr{\"a}henb{\"u}hl}, {Krause}, {La Barbera}, {Lelas}, {Leonardo},
  {Lewandowska}, {Lindfors}, {Lombardi}, {L{\'o}pez}, {L{\'o}pez-Coto},
  {L{\'o}pez-Oramas}, {Lorenz}, {Makariev}, {Maneva}, {Mankuzhiyil},
  {Mannheim}, {Maraschi}, {Mariotti}, {Mart{\'\i}nez}, {Mazin}, {Meucci},
  {Miranda}, {Mirzoyan}, {Mold{\'o}n}, {Moralejo}, {Munar-Adrover},
  {Niedzwiecki}, {Nieto}, {Nilsson}, {Nowak}, {Orito}, {Paiano}, {Paneque},
  {Paoletti}, {Pardo}, {Paredes}, {Partini}, {Perez-Torres}, {Persic}, {Pilia},
  {Pochon}, {Prada}, {Prada Moroni}, {Prandini}, {Puerto Gimenez}, {Puljak},
  {Reichardt}, {Reinthal}, {Rhode}, {Rib{\'o}}, {Rico}, {R{\"u}gamer},
  {Saggion}, {Saito}, {Saito}, {Salvati}, {Satalecka}, {Scalzotto}, {Scapin},
  {Schultz}, {Schweizer}, {Shore}, {Sillanp{\"a}{\"a}}, {Sitarek}, {Snidaric},
  {Sobczynska}, {Spanier}, {Spiro}, {Stamatescu}, {Stamerra}, {Steinke},
  {Storz}, {Strah}, {Sun}, {Suri{\'c}}, {Takalo}, {Takami}, {Tavecchio},
  {Temnikov}, {Terzi{\'c}}, {Tescaro}, {Teshima}, {Tibolla}, {Torres},
  {Treves}, {Uellenbeck}, {Vogler}, {Wagner}, {Weitzel}, {Zabalza}, {Zandanel},
  \& {Zanin}}]{Aleksic12}
{Aleksi{\'c}}, J., {Alvarez}, E.~A., {Antonelli}, L.~A., {et~al.} 2012, \aap,
  541, A13

\bibitem[{{Arikawa} {et~al.}(1999){Arikawa}, {Tatematsu}, {Sekimoto}, {Aso},
  {Noguchi}, {Shi}, {Miyazawa}, {Yamamoto}, {Ikeda}, {Maezawa}, {Ito}, {Saito},
  {Saito}, {Ozeki}, {Fujiwara}, {Inatani}, \& {Ohishi}}]{Arikawa99}
{Arikawa}, Y., {Tatematsu}, K., {Sekimoto}, Y., {et~al.} 1999, in Star
  Formation 1999, ed. T.~{Nakamoto}, 88--89

\bibitem[{{Bisbas} {et~al.}(2015){Bisbas}, {Papadopoulos}, \&
  {Viti}}]{Bisbas15}
{Bisbas}, T.~G., {Papadopoulos}, P.~P., \& {Viti}, S. 2015, \apj, 803, 37

\bibitem[{{Bisbas} {et~al.}(2017){Bisbas}, {van Dishoeck}, {Papadopoulos},
  {Sz{\H u}cs}, {Bialy}, \& {Zhang}}]{Bisbas17}
{Bisbas}, T.~G., {van Dishoeck}, E.~F., {Papadopoulos}, P.~P., {et~al.} 2017,
  \apj, 839, 90

\bibitem[{{Ceccarelli} {et~al.}(2011){Ceccarelli}, {Hily-Blant}, {Montmerle},
  {Dubus}, {Gallant}, \& {Fiasson}}]{Ceccarelli11}
{Ceccarelli}, C., {Hily-Blant}, P., {Montmerle}, T., {et~al.} 2011, \apjl, 740,
  L4

\bibitem[{{Ezawa} {et~al.}(2004){Ezawa}, {Kawabe}, {Kohno}, \&
  {Yamamoto}}]{Ezawa04}
{Ezawa}, H., {Kawabe}, R., {Kohno}, K., \& {Yamamoto}, S. 2004, in Society of
  Photo-Optical Instrumentation Engineers (SPIE) Conference Series, Vol. 5489,
  Ground-based Telescopes, ed. J.~{Oschmann}, Jacobus~M., 763--772

\bibitem[{{Frerking} {et~al.}(1982){Frerking}, {Langer}, \&
  {Wilson}}]{Frerking82}
{Frerking}, M.~A., {Langer}, W.~D., \& {Wilson}, R.~W. 1982, \apj, 262, 590

\bibitem[{{Furuya} \& {Aikawa}(2014)}]{furuya14}
{Furuya}, K., \& {Aikawa}, Y. 2014, \apj, 790, 97

\bibitem[{{Furuya} {et~al.}(2022){Furuya}, {Tsukagoshi}, {Qi}, {Nomura},
  {Cleeves}, {Lee}, \& {Yoshida}}]{furuya22}
{Furuya}, K., {Tsukagoshi}, T., {Qi}, C., {et~al.} 2022, \apj, 926, 148

\bibitem[{{Gabici} {et~al.}(2009){Gabici}, {Aharonian}, \&
  {Casanova}}]{Gabici09}
{Gabici}, S., {Aharonian}, F.~A., \& {Casanova}, S. 2009, \mnras, 396, 1629

\bibitem[{{Hocuk} {et~al.}(2017){Hocuk}, {Sz{\H{u}}cs}, {Caselli}, {Cazaux},
  {Spaans}, \& {Esplugues}}]{Hocuk17}
{Hocuk}, S., {Sz{\H{u}}cs}, L., {Caselli}, P., {et~al.} 2017, \aap, 604, A58

\bibitem[{{Iguchi} \& {Okuda}(2008)}]{Iguchi08}
{Iguchi}, S., \& {Okuda}, T. 2008, \pasj, 60, 857

\bibitem[{{Ikeda} {et~al.}(2002){Ikeda}, {Oka}, {Tatematsu}, {Sekimoto}, \&
  {Yamamoto}}]{Ikeda02}
{Ikeda}, M., {Oka}, T., {Tatematsu}, K., {Sekimoto}, Y., \& {Yamamoto}, S.
  2002, \apjs, 139, 467

\bibitem[{{Izumi} {et~al.}(2021){Izumi}, {Fukui}, {Tachihara}, {Fujita},
  {Torii}, {Kamazaki}, {Kaneko}, {Silva}, {Iono}, {Momose}, {Sugimoto},
  {Nakazato}, {Kosugi}, {Maekawa}, {Takahashi}, {Yoshino}, \&
  {Asayama}}]{Izumi21}
{Izumi}, N., {Fukui}, Y., {Tachihara}, K., {et~al.} 2021, \pasj, 73, 174

\bibitem[{{Kamegai} {et~al.}(2003){Kamegai}, {Ikeda}, {Maezawa}, {Ito},
  {Iwata}, {Sakai}, {Oka}, {Yamamoto}, {Sekimoto}, {Tatematsu}, {Noguchi},
  {Saito}, {Fujiwara}, {Ozeki}, {Inatani}, \& {Ohishi}}]{Kamegai03}
{Kamegai}, K., {Ikeda}, M., {Maezawa}, H., {et~al.} 2003, \apj, 589, 378

\bibitem[{{Koo} {et~al.}(1995){Koo}, {Kim}, \& {Seward}}]{Koo95}
{Koo}, B.-C., {Kim}, K.-T., \& {Seward}, F.~D. 1995, \apj, 447, 211

\bibitem[{Kuno {et~al.}(2011)Kuno, Takano, Iono, Nakajima, Iwashita, Handa,
  Hatsukade, Higuchi, Hirota, Ishikawa, Kaneko, Kawaguchi, Kawabe, Kimura,
  Kohno, Maekawa, Mikoshiba, Miyazawa, Miyazawa, Muraoka, Ogawa, Onodera,
  Saito, Takahashi, \& Yonezu}]{Kuno11}
Kuno, N., Takano, S., Iono, D., {et~al.} 2011, in 2011 XXXth URSI General
  Assembly and Scientific Symposium, 1--4

\bibitem[{{Massaro} {et~al.}(2015){Massaro}, {D'Abrusco}, {Landoni}, {Paggi},
  {Masetti}, {Giroletti}, {Ot{\'\i}-Floranes}, {Chavushyan},
  {Jim{\'e}nez-Bail{\'o}n}, {Pati{\~n}o-{\'A}lvarez}, {Digel}, {Smith}, \&
  {Tosti}}]{Massaro15}
{Massaro}, F., {D'Abrusco}, R., {Landoni}, M., {et~al.} 2015, \apjs, 217, 2

\bibitem[{{Minamidani} {et~al.}(2016){Minamidani}, {Umemoto}, {Nishimura},
  {Matsuo}, {Fujita}, {Tsuda}, {Ohashi}, {Tosaki}, \& {Kuno}}]{Minamidani16a}
{Minamidani}, T., {Umemoto}, T., {Nishimura}, A., {et~al.} 2016, in EAS
  Publications Series, Vol.~75, EAS Publications Series, 193--194

\bibitem[{{Oka} {et~al.}(2001){Oka}, {Yamamoto}, {Iwata}, {Maezawa}, {Ikeda},
  {Ito}, {Kamegai}, {Sakai}, {Sekimoto}, {Tatematsu}, {Arikawa}, {Aso},
  {Noguchi}, {Shi}, {Miyazawa}, {Saito}, {Ozeki}, {Fujiwara}, {Ohishi}, \&
  {Inatani}}]{Oka01}
{Oka}, T., {Yamamoto}, S., {Iwata}, M., {et~al.} 2001, \apj, 558, 176

\bibitem[{{Okuda} \& {Iguchi}(2008)}]{Okuda08}
{Okuda}, T., \& {Iguchi}, S. 2008, \pasj, 60, 315

\bibitem[{{Parsons} {et~al.}(2012){Parsons}, {Thompson}, {Clark}, \&
  {Chrysostomou}}]{Parsons12}
{Parsons}, H., {Thompson}, M.~A., {Clark}, J.~S., \& {Chrysostomou}, A. 2012,
  \mnras, 424, 1658

\bibitem[{{Sato} {et~al.}(2010){Sato}, {Reid}, {Brunthaler}, \&
  {Menten}}]{Sato10}
{Sato}, M., {Reid}, M.~J., {Brunthaler}, A., \& {Menten}, K.~M. 2010, \apj,
  720, 1055

\bibitem[{{Satou} {et~al.}(2008){Satou}, {Sekimoto}, {Iizuka}, {Ito}, {Shan},
  {Kamba}, {Kumagai}, {Kamikura}, {Tomimura}, {Serizawa}, {Asayama}, \&
  {Sugimoto}}]{Satou08}
{Satou}, N., {Sekimoto}, Y., {Iizuka}, Y., {et~al.} 2008, \pasj, 60, 1199

\bibitem[{{Shen} {et~al.}(2004){Shen}, {Greenberg}, {Schutte}, \& {van
  Dishoeck}}]{Shen04}
{Shen}, C.~J., {Greenberg}, J.~M., {Schutte}, W.~A., \& {van Dishoeck}, E.~F.
  2004, \aap, 415, 203

\bibitem[{{Shimajiri} {et~al.}(2013){Shimajiri}, {Sakai}, {Tsukagoshi},
  {Kitamura}, {Momose}, {Saito}, {Oshima}, {Kohno}, \& {Kawabe}}]{Shimajiri13}
{Shimajiri}, Y., {Sakai}, T., {Tsukagoshi}, T., {et~al.} 2013, \apjl, 774, L20

\bibitem[{{Tanaka} {et~al.}(2021){Tanaka}, {Nagai}, \& {Kamegai}}]{Tanaka21}
{Tanaka}, K., {Nagai}, M., \& {Kamegai}, K. 2021, \apj, 915, 79

\bibitem[{{Tielens}(2005)}]{Tielens05}
{Tielens}, A.~G.~G.~M. 2005, {The Physics and Chemistry of the Interstellar
  Medium} (Cambridge University Press)

\bibitem[{{Tielens} \& {Hollenbach}(1985)}]{Tielens85a}
{Tielens}, A.~G.~G.~M., \& {Hollenbach}, D. 1985, \apj, 291, 722

\bibitem[{{Vaupr{\'e}} {et~al.}(2014){Vaupr{\'e}}, {Hily-Blant}, {Ceccarelli},
  {Dubus}, {Gabici}, \& {Montmerle}}]{Vaupre14}
{Vaupr{\'e}}, S., {Hily-Blant}, P., {Ceccarelli}, C., {et~al.} 2014, \aap, 568,
  A50

\end{thebibliography}
\bibliographystyle{aasjournal}

\end{document}